\documentclass[sigplan,screen,nonacm]{acmart}

\usepackage{marginnote}

\usepackage{gdefs}
\usepackage{multirow}

\usepackage{thmtools}
\usepackage{thm-restate}
\usepackage[T1]{fontenc}

\usepackage{amsmath}
\usepackage{amsfonts}
\usepackage{graphicx}
\usepackage{setspace}
\usepackage{hhline}
\usepackage{refs}
\usepackage{multirow}
\usepackage{multicol}
\usepackage{wrapfig}
\usepackage{relsize}
\usepackage{stmaryrd}
\usepackage{galois}
\usepackage{kbordermatrix}
\usepackage{mdwlist} 
\usepackage{subcaption}

\usepackage[noline,noend,ruled,linesnumbered]{algorithm2e} 
\usepackage{color} 
\usepackage{mathpartir} 

\definecolor{light-gray}{gray}{0.9}

\usepackage[normalem]{ulem}

\usepackage{eqbox}

\usepackage{listings}
\usepackage{enumitem}
\setlist{nosep,leftmargin=*}

\usepackage{microtype}

\usepackage[leftbars,color]{changebar}

\newcommand{\revision}[1]{}

\cbcolor{red}

\newcommand{\subsubsubsection}[1]{\smallskip\noindent{\bf #1}}

\newcounter{LineNumber}[figure]

\newcommand{\AllExperiments}[1]{}

\newcommand{\Omit}[1]{}

\newcommand{\OnlyPaper}[1]{}
\newcommand{\OnlyTech}[1]{#1}

\newcommand{\sem}[1]{\llbracket #1 \rrbracket}
\newcommand{\relfunsem}[1]{\mathcal{F}\sem{#1}}
\newcommand{\relfunsemrec}[1]{\mathcal{F}_{\textit{rec}}\sem{#1}}
\newcommand{\relsem}[1]{\mathcal{R}\sem{#1}}
\newcommand{\expsem}[1]{\mathcal{E}\sem{#1}}
\newcommand{\basesem}[1]{\mathcal{F}_{\textit{base}}\sem{#1}}

\newcommand{\defeq}{\overset{\underset{\text{def}}{}}{=}}

\newcommand{\Var}{{\it Var}}
\newcommand{\State}{{\it State}}

\renewcommand{\iff}{\operatorname{\Leftrightarrow}}

\newcommand{\tuple}[1]{(#1)}

\newcommand{\prog}[1]{\texttt{#1}}

\newcommand{\bsym}[2]{b_{#1}(#2)} 
\newcommand{\bsymsuper}[3]{b^{#3}_{#1}(#2)}

\newcommand{\bsymGeneral}[1]{b_{#1}} 

\newcommand{\summaryrec}{\varphi_{\textit{rec}}}
\newcommand{\summaryrecfor}[1]{\varphi_{\textit{rec}(#1)}}
\newcommand{\summarybase}{\beta}
\newcommand{\summarybasefor}[1]{\beta_{#1}}
\newcommand{\summaryfor}[1]{\varphi_{#1}}
\newcommand{\summarycall}{\varphi_{\textit{call}}}
\newcommand{\summarycallfor}[1]{\varphi_{\textit{call}(#1)}}
\newcommand{\summaryextract}{\varphi_{\textit{ext}}}
\newcommand{\summaryextractfor}[1]{\varphi_{\textit{ext}(#1)}}
\newcommand{\summarydepth}[1]{\zeta_{#1}} 
\newcommand{\summaryheight}{\varphi_{\textit{height}}}

\newcommand{\depthone}{\varphi_{[D:=1]}}
\newcommand{\depthinc}{\varphi_{[D:=D+1]}}
\newcommand{\ventry}[1]{e_{#1}}
\newcommand{\vexit}[1]{x_{#1}}
\newcommand{\sccentry}[1]{e_{#1}'}
\newcommand{\sccexit}{x'}
\newcommand{\weightededges}{E}
\newcommand{\calledges}{C}
\newcommand{\newedges}{E'}

\newcommand{\vars}{\textit{Var}}

\newcommand{\wedgebase}{w_{\textit{base}}}
\newcommand{\wedgeextract}[1]{w_{\textit{ext},#1}}

\newcommand{\boundsmap}{\textit{DefinesBound}}
\newcommand{\depsmap}{\textit{UsesBound}}
\newcommand{\nonlindepsmap}{\textit{UsesBoundNonLinearly}}

\newcommand{\Chora}{\textsf{C{\smaller HORA}}}
\newcommand{\benchmark}[1]{\textit{#1}}

\begin{document}

\title{Templates and Recurrences: Better Together}


\author{Jason Breck}
\email{jbreck@cs.wisc.edu}
\affiliation{%
	\institution{University of Wisconsin}
	\city{Madison}
	\state{WI}
	\country{USA}
}
\author{John Cyphert}
\email{jcyphert@wisc.edu}
\affiliation{%
  \institution{University of Wisconsin}
  \city{Madison}
  \state{WI}
  \country{USA}
}
\author{Zachary Kincaid}
\email{zkincaid@cs.princeton.edu}
\affiliation{%
  \institution{Princeton University}
  \city{Princeton}
  \state{NJ}
  \country{USA}
}
\author{Thomas Reps}
\email{reps@cs.wisc.edu}
\affiliation{%
  \institution{University of Wisconsin}
  \city{Madison}
  \state{WI}
  \country{USA}
}


\begin{abstract}


This paper is the confluence of two streams of ideas in the
literature on generating numerical invariants, namely:
(1) template-based methods, and (2) recurrence-based methods.

A \emph{template-based method} begins with a template that contains
unknown quantities, and finds invariants that match the template by
extracting and solving constraints on the unknowns.
A disadvantage of template-based methods is that they require fixing the set of
terms that may appear in an invariant in advance.  This disadvantage is
particularly prominent for non-linear invariant generation,
because the user must supply maximum degrees on polynomials, bases for exponents, etc.

On the other hand, \emph{recurrence-based methods} are able to find
sophisticated non-linear mathematical relations, including polynomials,
exponentials, and logarithms, because such relations arise as the solutions to
recurrences.  
However, a disadvantage of past recurrence-based invariant-generation methods
is that they are primarily loop-based analyses: they use recurrences to relate
the pre-state and post-state of a loop, so it is not obvious how to apply them
to a recursive procedure, especially if the procedure is \emph{non-linearly recursive}
(e.g., a tree-traversal algorithm).

In this paper, we combine these two approaches and obtain a technique that uses
templates in which the unknowns are \textit{functions} rather than numbers, and
the constraints on the unknowns are \textit{recurrences}.  The technique
synthesizes invariants involving polynomials, exponentials, and logarithms,
even in the presence of arbitrary control-flow, including any combination of
loops, branches, and (possibly non-linear) recursion. 
For instance, it is able to show that
(i) the time taken by merge-sort is $O(n \log(n))$, and
(ii) the time taken by Strassen's algorithm is $O(n^{\log_2(7)})$.
\OnlyTech{

This paper is an extended version of a paper with the same title at PLDI 2020
\cite{PLDI:BCKR2020}.
}
\end{abstract}



\maketitle

\setlength{\algomargin}{2.0ex}
\SetInd {0.0em}{0.7em} 


%
%
%

\section{Introduction}
\label{Se:Introduction}

%
%

A large body of work within the numerical-invariant-generation literature
focuses on \textit{template-based methods} \cite{CAV:CSS03,SAS:SSM04}.
Such methods fix the form of the invariants that can be discovered, by
specifying a template that contains unknown quantities.  Given a program and
some property to be proved, a template-based analyzer proceeds by finding
constraints on the values of the unknowns and then solving these constraints
to obtain invariants of the program that suffice to prove the property.
Template-based methods have been particularly successful for finding invariants
within the domain of linear arithmetic.

Many programs have important numerical invariants that involve
non-linear mathematical relationships, such as polynomials, exponentials, and
logarithms.  
A disadvantage of template-based methods for non-linear invariant generation
is that (in contrast to the linear case) there is no ``most general''
template term, so the user must supply the set of terms that may appear in
the invariant.

In this paper, we present an invariant-synthesis technique that is related
to template-based methods, but sidesteps the above difficulty.
Our technique is based on a concept that we call a 
\textit{hypothetical summary}, which is a template for a procedure summary in
which the unknowns are \textit{functions}, rather than numbers.
The constraints that we extract for these functions are \textit{recurrences}.
Solving these recurrence constraints allows us to synthesize terms over program
variables that we can substitute in place of the unknown functions in our
template and thereby obtain procedure summaries.

Whereas most template-based methods directly constrain the mathematical form of
their invariants, our technique constrains the invariants indirectly, by way of
recurrences, and thereby allows the invariants to have a wide variety of
mathematical forms involving polynomials, exponentials, and logarithms.
This aspect is intuitively illustrated by the recurrences $S(n) = 2S(n/2) + n$ and
$T(n) = 2T(n/2) + n^2$:
although these two recurrences are outwardly similar, their solutions are more
different than one would expect at first glance, 
in that $S(n)$ is $\Theta(n \log n)$, whereas $T(n)$ is $\Theta(n^2)$.
Because the unknowns in our templates are functions, we can generate a wide
variety of invariants (involving polynomials, exponentials, logarithms) without
specifying their exact syntactic form.

However, recurrence-based invariant-generation techniques typically have
disadvantages when applied to recursive programs.  Recurrences are well-suited
to characterize the sequence of states that occur as a loop executes.
This idea can be extended to handle \emph{linear recursion}---where a
recursive procedure makes only a \emph{single} recursive call:
each procedure-entry state that occurs ``on the way down'' to the base case of the
recursion is paired with the corresponding procedure-exit state that occurs ``on the way
back up'' from the base case, and then recurrences are used to describe the
sequence of such state pairs. 
However, \emph{non-linear recursion} has a different structure:
it is tree-shaped,
rather than linear, and thus some kind of additional abstraction is
required before non-linear recursion can be described using recurrences.

We use the technique of hypothetical summaries to extend the work of
\cite{FMCAD:FK15}, \cite{PACMPL:KCBR18}, and \cite{PLDI:KBFR17}:
hypothetical summaries enable a different approach to the analysis of
non-linearly recursive programs, such as divide-and-conquer or tree-traversal
algorithms.\footnote{
  Warning: We use the term ``non-linear'' in two different senses:
  \emph{non-linear recursion} and \emph{non-linear arithmetic}.
  Even for a loop that uses linear arithmetic, non-linear arithmetic may
  be required to state a loop invariant.
  Moreover, arithmetic expressions in the programs that we analyze are
  not limited to linear arithmetic: variables can be multiplied.

  \hspace*{1.5ex}
  The two uses of the term ``non-linear'' are essentially unrelated,
  and which term is intended should be clear from context.
  The paper primarily concerns new techniques for handling non-linear recursion,
  and non-linear arithmetic is handled by known methods, e.g., \cite{PACMPL:KCBR18}.
}
We show how to analyze the base case of a procedure to extract a template
for a procedure summary (i.e., a hypothetical summary).  By assuming that every
call to the procedure, throughout the tree of recursive calls, is consistent
with the template, we discover relationships (i.e., recurrence constraints)
among the states of the program at different heights in the tree.  We then
solve the constraints and fill in the template to obtain a procedure summary.
Hypothetical summaries thus provide the additional layer of abstraction that is
required to apply recurrence-based invariant generation to non-linearly
recursive procedures.

Our invariant generation procedure is both (1)
\textit{general-purpose}, so it is applicable to a wide variety of
tasks, and (2) \textit{compositional}, so the space and time required
to analyze a program fragment depends on the size of the fragment
rather than the whole program.  In contrast, conventional
template-based methods are goal-directed (they must be tailored to a
specific problem of interest, e.g., a template-based invariant
generator for verification problems cannot solve quantitative problems
such as resource-bound analysis) and whole-program.  The
general-purpose nature of our procedure also distinguishes it from
recurrence-based resource-bound analyses, which for example cannot be
applied to assertion checking.

To evaluate the applicability of our analysis to challenging
numerical-invariant-synthesis tasks,
we applied it to the task of generating
bounds on the computational complexity of non-linearly recursive programs
and the task of generating invariants that suffice to prove assertions.
Our experiments show that the analysis technique is able to prove properties
that \cite{PLDI:KBFR17} was not capable of proving, and is competitive
with the output of state-of-the-art assertion-checking and 
resource-bound-analysis tools.

\subsubsubsection{Contributions.}
Our work makes contributions in three main areas:
\begin{enumerate}
  \item
    We introduce an analysis method based on ``hypothetical summaries.''
    It hypothesizes that a summary exists of a particular form, using uninterpreted function symbols
    to stand for unknown expressions.
    Analysis is performed to obtain constraints on the function symbols, which are then
    solved to obtain a summary.
  \item
    We develop a procedure-summarization technique called height-based
    recurrence analysis, which uses the notion of hypothetical summaries to
    produce bounds on the values of program variables based on the height of
    recursion (\sectref{Height}). 
    \OnlyTech{We further develop algorithms that, when used in conjunction
    with height-based recurrence analysis (\sectrefs{DepthBound}{DualHeight}),
    yield more precise summaries.}
    Furthermore, we give an algorithm (\sectref{Mutual}) that generalizes
    height-based recurrence analysis to the setting of mutual recursion.
  \item
    The technique is implemented in the $\Chora$ tool.
    Our experiments show that $\Chora$ is able to handle many non-linearly recursive
    programs, and generate invariants that include exponentials, polynomials,
    and logarithms (\sectref{Experiments}).
    For instance, it is able to show that
    (i) the time taken by merge-sort is $O(n \log(n))$,
    (ii) the time taken by Strassen's algorithm is $O(n^{\log_2(7)})$, and
    (iii) an iterative function and a non-linearly recursive function
    that both perform exponentiation are functionally equivalent.
\end{enumerate}
\Omit{

\subsubsubsection{Organization.}}
\sectref{Overview} presents an example to provide intuition.
\sectref{Background} provides background on material needed for understanding
 the paper's results.
 \Omit{\sectref{TechnicalDetails} presents our method(s) for extracting a recurrence
 inequations that over-approximates the behavior of a program.
 \sectref{Experiments} presents experimental results.}
\sectref{Related} discusses related work.



\section{Overview}
\label{Se:Overview}

\begin{figure}[t]
	\[
	\begin{array}{l}
	\hspace{0.0cm} \textrm{int}~nTicks;~
	\textrm{bool}~found;                                               \\
	\hspace{0.0cm} \textrm{int}~subsetSum(\textrm{int} * A,                 
	\textrm{int}~n)~\{ \\
	\hspace{0.5cm} \textit{found}=\textrm{false};~
	\textbf{return}~subsetSumAux(A,0,n,0);          \\
	\hspace{0.0cm} \}           \\
	\hspace{0.0cm} \textrm{int}~subsetSumAux(\textrm{int} * A, \textrm{int}~i,  
	\textrm{int}~n, \textrm{int}~sum)~\{          \\
	\hspace{0.5cm}     nTicks\textrm{++};                                             \\
	\hspace{0.5cm}     \textbf{if}~(i >= n)~\{~                                       \\
	\hspace{1.0cm}         \textbf{if}~(sum == 0) ~\{~\textit{found}=\textrm{true};\} \\
	\hspace{1.0cm}         \textbf{return}~0;                                         \\
	\hspace{0.5cm}     \}                                                             \\
	\hspace{0.5cm}     \textrm{int}~size=subsetSumAux(A,i+1,n,sum+A[i]); \\ 
	\hspace{0.5cm}     \textbf{if}~(\textit{found})~\{~ \textbf{return}~size + 1; \}  \\
	\hspace{0.5cm}     size=subsetSumAux(A,i+1,n,sum); \\     
	\hspace{0.5cm}     \textbf{return}~size;                                          \\
	\hspace{0.0cm} \}                                                                 \\
	\end{array}
	\]
	
\includegraphics[width=0.48\textwidth]{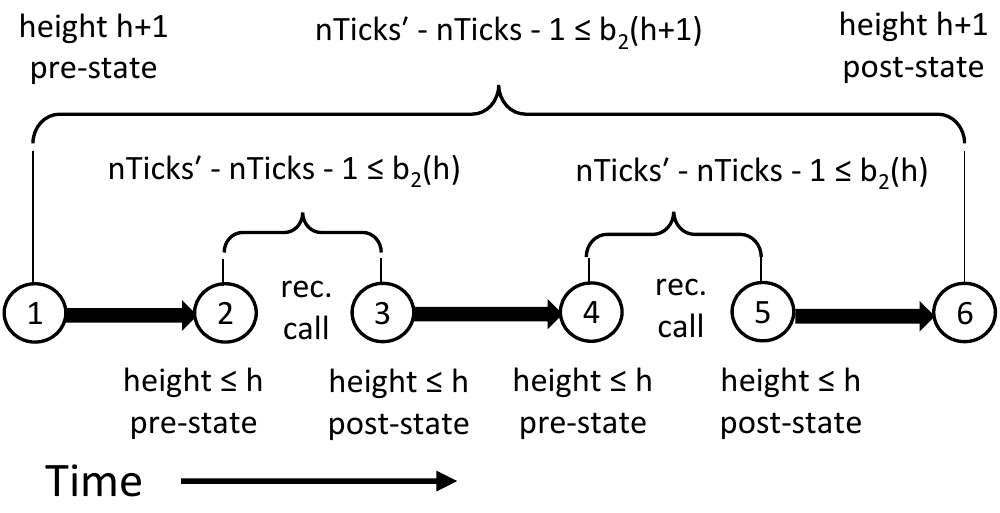}
\caption{\label{Fi:subsetSum}
{\small Example program \textit{subsetSum}. The diagram at the bottom shows a
timeline of a height $(h+1)$ execution of \textit{subsetSumAux}.
$\bsym{2}{h+1}$ is related to the 
increase of $\prog{nTicks}$ between the pre-state (label 
1) and the post-state (label 6).  
$\bsym{2}{h}$ is related to the increase of $\prog{nTicks}$ between (2) and (3)
and also between (4) and (5), i.e., between the pre-states and post-states
of height-$h$ executions.}
}
\end{figure}

The goal of this paper is to find numerical summaries for all
the procedures in a given program.  For simplicity, this section discusses
the analysis of a program that contains a single procedure $P$, which is
non-linearly recursive and calls no other procedures.

We use the following example to illustrate how our techniques use
recurrence solving to summarize non-linearly-recursive procedures.
\begin{example} \label{Exa:SubsetSum}
The function \textit{subsetSum} (\figref{subsetSum}) takes an array $\prog{A}$ of $\prog{n}$
integers, and performs a brute-force search to determine whether any non-empty
subset of $\prog{A}$'s elements sums to zero.  If it finds such a set, it
returns the number of elements in the set, and otherwise it returns zero.
The recursive function \textit{subsetSumAux} works by sweeping through the
array from left to right, making two recursive calls for each array element.
The first call considers subsets that include the element $\prog{A[i]}$, and the
second call considers subsets that exclude $\prog{A[i]}$.  The sum of the
values in each subset is computed in the accumulating parameter $\prog{sum}$.
When the base case is reached, \textit{subsetSumAux} checks whether
$\prog{sum}$ is zero, and if so, sets $\prog{found}$ to \textit{true}.  
At each of the two recursive call sites, the value returned by the recursive
call is stored in the variable $\prog{size}$.  After $\prog{found}$ is set to
\textit{true}, \textit{subsetSumAux} computes the size of the subset by
returning $\prog{size} + 1$ if the subset was found after the first recursive
call, or returning $\prog{size}$ unchanged if the subset was found after the
second recursive call.

In this paper, a \textbf{state} of a program is an assignment of integers to
program variables.  
For each procedure $P$, we wish to characterize the relational semantics $R(P)$,
defined as the set of state pairs $(\sigma,\sigma')$ such that $P$ can start
executing in state $\sigma$ and finish in state $\sigma'$.
To find an over-approximate representation of the relational semantics of a
recursive procedure such as \textit{subsetSumAux}, we take an approach that we
call \emph{height-based recurrence analysis}.
In height-based recurrence analysis, we construct and solve recurrence relations
to discover properties of the transition relation of a recursive procedure.
To formalize our use of recurrence relations, we give the following definitions.

We define the \textit{height-bounded relational semantics} $R(P,h)$ to be the
subset of $R(P)$ that $P$ can achieve if it is limited to using an execution
stack with a height of at most $h$ activation records.  We define a
height-$h$ execution of $P$ to be any execution of $P$ that uses a
stack height of at most $h$, or, in other words, an execution of $P$ having
recursion depth no more than $h$. 
Base cases are defined to be of height 1.
Let $\tau_1,...,\tau_n$ be a set of polynomials over unprimed and primed program
variables, representing the pre-state and post-state of $P$, respectively.
For each $\tau_k$ we associate a function $V_k :\mathbb{N}\rightarrow 2^{\mathbb{Q}}$, such that $V_k(h)$ is defined to
be the set of values $v$ such that, for some $(\sigma,\sigma') \in R(P,h)$,
$\tau_k$ evaluates to $v$ by using $\sigma$ and $\sigma'$ to interpret the unprimed and primed variables, respectively.

Using \textit{subsetSumAux} as an example, let $\tau_1 \defeq \prog{return}'$. Then, $V_1(1)$ denotes the set of values $\prog{return}'$ can take on in \emph{any} base case of \textit{subsetSumAux}. In this program, $\prog{return}'$ is 0 in any base case, and so $V_1(1) = \{0\}$. Now consider an execution of height 2. In the case that $\prog{found}$ is true, we have that $\prog{return}'$ increases by 1 compared to the value that $\prog{return}'$ has in the base case. If $\prog{found}$ is not true then $\prog{return}'$ remains the same. In other words, at height-2 executions, $\prog{return}'$ takes on the values 0 and 1; i.e., $V_1(2) = \{0, 1\}$. Similarly, $V_1(3) = \{0, 1, 2\}$, and so on.
We approximate the value set $V_k(h)$ by finding a function $b_k(h):\mathbb{N}\rightarrow \mathbb{Q}$ that bounds $V_k(h)$ for all $h$; that is, for any $v\in V_k(h)$, we have $v\le b_k(h)$. In the case of $\tau_1$, a suitable bounding function $b_1(h)$ is $b_1(h) = h-1$. The initial step of our analysis chooses terms $\tau_1, ..., \tau_n$, and then for each term $\tau_k$, tries to synthesize a function $b_k(h)$ that bounds the set of values $\tau_k$ can take on.

Note that for a given term $\tau_j$, a corresponding bounding function may not exist. A necessary condition for a bounding function to exist for a term $\tau_j$ is that the set $V_j(1)$ must be bounded. This observation restricts our set of candidate terms $\tau_1,...,\tau_n$ to only be over terms that are bounded above in the base case. (Specifically, we require the expressions to be bounded above by
zero.) For example, $\prog{return}'\le 0$ in the base case, and so $\tau_1 \defeq \prog{return}'$ is a candidate term. Similarly, the term $\tau_2\defeq \prog{nTicks'-nTicks-1}$ is also bounded above by 0 in the base case, and so $\tau_2$ is a candidate term. There are other candidate terms that our analysis would extract for this example, but for brevity they are not listed here. We discover these bounded terms $\tau_1$ and $\tau_2$ using \emph{symbolic abstraction} (see \sectref{Background}).

Once we have a set of candidate terms $\tau_1,...,\tau_n$, we seek to find corresponding bounding functions $b_1(h), ..., b_k(h)$. Note that such functions may not exist: just because $\tau_k$ is bounded above in the base case does not mean it is bounded in all other executions.  If a bounding function for a term does exist, we would like a closed-form expression for it in terms of $h$. We derive such closed-form expressions by \emph{hypothesizing} that a bounding function $b_k(h)$ does exist. These hypothetical functions $b_k(h)$ allow us to construct a \emph{hypothetical procedure summary} $\varphi_h$ that represents a
typical height-$h$ execution. For example, in the case of \textit{subsetSumAux}:
\[
\varphi_{h}
~\eqdef~
\prog{return}' \leq \bsym{1}{h}
~\land~
\prog{nTicks}' - \prog{nTicks} - 1 \leq \bsym{2}{h}.
\]
Note that, although
$\varphi_h$ assumes the existence of several bounding functions 
(corresponding to $\bsym{k}{h}$ for several values of $k$), the assumptions
for different values of $k$ need not all succeed or fail together.  That is, if we fail
to find a bounding function $\bsym{k}{h}$ for some $k$, this failure does not prevent us
from continuing the analysis and finding other bounding functions
($\bsym{j}{h}$, with $j\neq k$) for the same procedure.

We then build up a height-$(h+1)$ summary, $\varphi_{h+1}$, compositionally, with $\varphi_{h}$ replacing the recursive calls. For example, consider the term $\tau_2 = \prog{nTicks}' - \prog{nTicks} - 1$ in the context of \figref{subsetSum}. Our goal is to create a relational summary for the variable $\prog{nTicks}$ between labels 1 and 6. We do this by extending a summary for the transition between labels 1 and 2 with a summary for the transition between 2 and 3, namely, our hypothetical summary. Then we extend that with a summary for the paths between labels 3 and 4, and so on. Between labels 1 and 2, $\prog{nTicks}$ gets increased by 1.
We then summarize the transition between 1 and 3. We know $\prog{nTicks}$ gets increased by 1 between labels 1 and 2. Furthermore, our hypothetical bounding function $\prog{nTicks}' - \prog{nTicks} - 1 \leq \bsym{2}{h}$ says that $\prog{nTicks}$ gets increased by at most $\bsym{2}{h} + 1$ between labels 2 and 3. Combining these summaries, we see that $\prog{nTicks}$ gets increased by at most $\bsym{2}{h} + 2$ between labels 1 and 3. $\prog{nTicks}$ does not change between labels 3 and 4, so the summary between labels 1 and 4 is the same as the one between labels 1 and 3. The transition between labels 4 and 5 is a recursive call, so we again use our hypothetical summary to approximate this transition. Once again, such a summary says $\prog{nTicks}$ gets increased by at most $\bsym{2}{h} + 1$. Extending our summary for the transition between 1 and 4 with this information allows us to conclude that $\prog{nTicks}$ gets increased by at most $2\bsym{2}{h}+3$ between labels 1 and 5. $\prog{nTicks}$ does not change between labels 5 and 6. Consequently, our summary for $\prog{nTicks}$ between labels 1 and 6 is $\prog{nTicks}' - \prog{nTicks} \le 2\bsym{2}{h}+3$. Similar reasoning would also obtain a summary for $\prog{return}$ as $\prog{return}' \le 1 + \bsym{1}{h}$. These formulas constitute our height-$(h+1)$ hypothetical summary, $\varphi_{h+1}$.
\[
  \varphi_{h+1}
~\eqdef~
  \prog{return}' \leq 1 + \bsym{1}{h}
    ~\land~
  \prog{nTicks}' \leq \prog{nTicks} + 2\bsym{2}{h} + 3
\]
If we rearrange each conjunct to respectively place $\tau_1$ and $\tau_2$ on the left-hand-side of each inequality, we obtain height-$(h+1)$ bounds on the values of $\tau_1$ and $\tau_2$. By definition such bounds are valid expressions for $\bsym{1}{h+1}$ and $\bsym{2}{h+1}$. That is at height-$(h+1)$,
\begin{align}
   \prog{return}' &\leq \bsym{1}{h} + 1 = \bsym{1}{h+1}\label{Eq:returnRec} \\
  \prog{nTicks}' - \prog{nTicks} - 1 &\leq  2 + 2\bsym{2}{h} = \bsym{2}{h+1}\label{Eq:nTicksRec}
\end{align}
The equations give recursive definitions for $\bsymGeneral{1}$ and $\bsymGeneral{2}$. Solving these recurrence relations give us bounds on the value sets $V_1(h)$ and $V_2(h)$, for all heights $h$.

In \sectref{DepthBound}, we present an algorithm that determines an upper
bound on a procedure's depth of recursion as a function of the parameters to
the initial call and the values of global variables.  This depth of recursion
can also be interpreted as a stack height $h$ that we can use as an argument
to the bounding functions $\bsym{k}{h}$.  In the case of \textit{subsetSumAux},
we obtain the bound 
$h \leq \max(1,1+\prog{n}-\prog{i})$.
The solutions to the recurrences discussed above, when combined with the depth
bound, yield the following summary.
\begin{align*}
\prog{nTicks}' &\leq \prog{nTicks} + 2^{h} -1
~\land~
\prog{return}' \leq h - 1 
~\land~\\
h &\leq \max(1,1+\prog{n} - \prog{i})
\end{align*}
When $\textit{subsetSum}$ is called with some array size $n$, the maximum
possible depth of recursion that can be reached by $\textit{subsetSumAux}$
is equal to $n$.  In this way, we have established that the running time of
$\textit{subsetSum}$ is exponential in $n$, and the return value is at most
$n$.
\end{example}



\section{Background}
\label{Se:Background}

\paragraph{Relational semantics.}
In the following, we give an abstract presentation of the relational
semantics of programs.  Fix a set \Var{} of program variables.  A
\textbf{state} $\sigma : \State{} \eqdef \Var{} \rightarrow
\mathbb{Z}$ consist of an integer valuation for each program variable.
A recursive procedure $P$ can be understood as a chain-continuous (and
hence monotonic) function on state relations $\relfunsem{P} :
2^{\State{} \times \State{}} \rightarrow 2^{\State{} \times
  \State{}}$.  The \textbf{relational semantics} $\relsem{P}$ of $P$
is given as the limit of the ascending Kleene chain of
$\relfunsem{P}$:
  \begin{align*}
    R(P,0) &= \emptyset\\
    R(P,h+1) &= \relfunsem{P}(R(P,h))\\
    \relsem{P} &= \bigcup_{h \in \mathbb{N}} R(P,h)
  \end{align*}
  Operationally, for any $h$ we may view $R(P,h)$ as the input/output
  relation of $P$ on a machine with a stack limit of $h$ activation
  records.  We can extend relational semantics to mutually recursive
  procedures in the natural way, by considering $\relfunsem{P}$ to be
  function that takes as input a $k$-tuple of state relations (where
  $k$ is the number of mutually recursive procedures).

  A \textbf{transition formula} $\varphi$ is a formula over the program
  variables $\Var{}$ and an additional set $\Var{}'$ of ``primed''
  copies, representing the values of the program variables before and
  after a computation.  A transition relation $\varphi$ can be
  interpreted as a property that holds of a pair of states
  $\tuple{\sigma,\sigma'}$: we say that $\tuple{\sigma,\sigma'}$
  satisfies $\varphi$ if $\varphi$ is true when each variable in $\Var{}$ is
  interpreted according to $\sigma$, and each variable in $\Var{}'$ is
  interpreted according to $\sigma'$.  We use $\relsem{\varphi}$ to denote
  the state relation consisting of all pairs $\tuple{\sigma,\sigma'}$
  that satisfy $\varphi$.  This paper is concerned with the problem of
  \textit{procedure summarization}, in which the goal is to find a
  transition formula $\varphi$ that \textit{over-approximates} a
  procedure, in the sense that $\relsem{P} \subseteq \relsem{\varphi}$.

  A \textbf{relational expression} $\tau$ is a polynomial over $\Var{}
  \cup \Var{}'$ with rational coefficients.  A relational expression
  can be evaluated at a state pair $\tuple{\sigma,\sigma'} \in \State
  \times \State$ by using $\sigma$ to interpret the unprimed symbols
  and $\sigma'$ to interpret the primed symbols---we use
  $\expsem{\tau}(\sigma,\sigma')$ to denote the evaluation of $\tau$
  at $\tuple{\sigma,\sigma'}$.

\paragraph{Intra-procedural analysis.}
  The technique for procedure summarization developed in this paper makes
  use of \textit{intra}-procedural summarization as a sub-routine.  We
  formalize this intra-procedural technique by a function
  $\textit{PathSummary}(e,x,V,E)$, which takes as input a control-flow
  graph with vertices $V$, edges $E$, entry vertex $e$, and exit
  vertex $x$, and computes a transition formula that over-approximates
  all paths in $(V,E)$ between $e$ and $x$. We use
  $\textit{Summary}(P,\varphi)$ to denote a function that takes as
  input a recursive procedure $P$ and a transition formula $\varphi$,
  and computes a transition formula that over-approximates $P$ when
  $\varphi$ is used to interpret recursive calls (i.e.,
  $\relfunsem{P}(\relsem{\varphi}) \subseteq
  \relsem{\textit{Summary}(P,\varphi)}$).
  $\textit{Summary}(P,\varphi)$ can be implemented in terms of
  $\textit{PathSummary}(e,x,V,E)$ by replacing all call edges with
  $\varphi$, and taking $(e,x,V,E)$ to be the control-flow graph of
  $P$.

  In principle, any intra-procedural summarization procedure can be
  used to implement $\textit{Summary}(P,\varphi)$; the implementation
  of our method uses the technique from \citet{PACMPL:KCBR18}.

\begin{algorithm}[tb]
{\small
  \SetKwInOut{Input}{Input}
  \SetKwInOut{Output}{Output}
  \Input{Formula of the form $\exists X. \psi$ where $\psi$ is satisfiable and quantifier-free}
  \Output{Convex hull of $\exists X. \psi$}
  $P \gets \bot$\;
  \While{\textit{there exists a model} $m$ \textit{of} $\psi$}{
     Let $Q$ be a cube of the DNF of $\psi$ s.t. $m \models Q$\;
     $Q \gets \textit{project}(Q,X)$ \tcc*{Polyhedral projection}
     $P \gets P \sqcup Q$ \tcc*{Polyhedral join}
     $\psi \gets \psi \land \lnot P$\;
  }
  \Return{$P$}
}
\caption{The convex-hull algorithm from \cite{FMCAD:FK15}\label{Alg:PolyhedralAlphaHat}}
\end{algorithm}

\paragraph{Symbolic abstraction.}
We use $\textit{Abstract}(\varphi,V)$ to denote a
procedure that takes a formula $\varphi$ and computes a set of
polynomial inequations over the variables $V$ that are implied by
$\varphi$.  
If $\varphi$ is expressed in linear arithmetic, then a representation of
\textit{all} implied polynomial inequations (namely, a constraint
representation of the convex hull of $\varphi$ projected onto $V$) can be
computed effectively (e.g., using \cite[Alg.~2]{FMCAD:FK15},
which we show in this paper as \algref{PolyhedralAlphaHat}).  
Otherwise, we settle for a \textit{sound} procedure that produces inequations
implied by $\varphi$, but not necessarily all of them (e.g., using
\cite[Alg.~3]{PACMPL:KCBR18}).

In principle, the convex hull of a linear arithmetic formula F can be computed
as follows: write F in disjunctive normal form, as $F \equiv C_1 \lor ... \lor
C_n$, where each $C_i$ is a conjunction of linear inequations (i.e., a convex
polyhedron).  The convex hull of $F$ is obtained by replacing disjunctions with
the join operator of the domain of convex polyhedra.  This algorithm can be
improved by using an SMT solver to enumerate the DNF lazily, and extended to
handle existential quantification by using polyhedral projection 
(\algref{PolyhedralAlphaHat}).  
A similar approach can be used to compute a conjunction of non-linear
inequations that are implied by a formula $F$, by treating non-linear terms in
the formula as additional dimensions of the space (e.g., a quadratic inequation
$x^2 < y^2$ is treated as a linear inequation $d_{x^2} < d_{y^2}$, where
$d_{x^2}$ and $d_{y^2}$ are symbols that we associate with the terms $x^2$ and
$y^2$, but have no intrinsic meaning).  The non-linear variation of the
algorithm's precision can be improved by using inference rules, congruence
closure, and Grobner-basis algorithms to deduce linear relations among the
non-linear dimensions that are consequences of the non-linear theory
(\cite[Alg.~3]{PACMPL:KCBR18}).  Note that, because non-linear integer
arithmetic is undecidable, this process is (necessarily) incomplete.

\paragraph{Recurrence relations.}
\emph{C-finite sequences} are a well-studied class of sequences
defined by linear recurrence relations, of which a famous example is
the Fibonacci sequence.  Formally,
\begin{definition}\label{De:cFinite}
  A sequence $s : \mathbb{N} \rightarrow \mathbb{Q}$ is $C$-finite of order $d$ if it satisfies a linear recurrence equation
\[ s(k+d) = c_1s(k+d-1)+...+c_{d-1}s(k+1)+c_ds(k)\ ,\] where each $c_i$ is a constant.
\end{definition}
It is classically known that every C-finite sequence $s(k)$ admits a closed form that is computable from its recurrence relation and takes the form of an exponential-polynomial
\[
s(k) = p_1(k)r_1^k + p_2(k)r_2^k+...+p_l(k)r_l^k\ ,
\]
where each $p_i$ is a polynomial in $k$ and each $r_i$ is a constant.
In the following, it will be convenient to use a different kind of
recurrence relation to present $C$-finite sequences, namely
\textit{stratified systems of polynomial recurrences}. 
\begin{definition} \label{De:solvable}
  A \textit{stratified system of polynomial recurrences} is a system
  of recurrence equations over sequences $x_{1,1},...,x_{1,n_1},...,x_{m,1},...,x_{m,n_m}$
  of the form
  \[ \{ x_{i,j}(k+1) = c_{i,j,1}x_{i,1}(k) +
  \dotsi + c_{i,j,n_i}x_{i,n_i}(k) + p_{i,j} \}_{i,j} \]
  where each $c_{i,j,1}, ..., c_{i,j,n_i}$ is a constant, and
  $p_{i,j}$ is a polynomial in
  $x_{1,1}(k),...,x_{1,n_1}(k),...,x_{i-1,1}(k),...,x_{i-1,n_{i-1}}(k)$.
\end{definition}

Intuitively, the sequences
$x_{1,1},...,x_{1,n_1},...,x_{m,1},...,x_{m,n_m}$ are organized into
\textit{strata} ($x_{1,1},...,x_{1,n_1}$ is the first,
$x_{2,1},...,x_{2,n_1}$ is the second, and so on), the right-hand-side
of the equation for $x_{i,j}$ can involve \textit{linear} terms over
the sequences in the $i^{\textit{th}}$ strata, and additional
\textit{polynomial} terms over sequences of lower strata.  It follows
from the closure properties of C-finite sequences that each $x_{i,j}$
defines a $C$-finite sequence, and an exponential-polynomial closed
form for each sequence can be computed from a stratified system of
polynomial recurrences \cite{Book:KP2011}.  The fact that \textit{any}
$C$-finite sequence satisfies a stratified system of polynomial
recurrences follows from the fact that a recurrence of order $d$ can
be implemented as a system of \textit{linear} recurrences among $d$
sequences \cite{Book:KP2011}.

\begin{example}
  An example of a stratified system of polynomial recurrences with
  four sequences ($w,x,y,z$) arranged into two strata ($(w,x)$ and
  $(y,z)$) is as follows:
  \begin{align*}
    \begin{bmatrix}
      w(k+1)\\
      x(k+1)
    \end{bmatrix}
    &=
    \begin{bmatrix}
      1 & \frac{1}{3} \\
      0 & 2
    \end{bmatrix}
    \begin{bmatrix}
      w(k)\\
      x(k)
    \end{bmatrix}
    +
    \begin{bmatrix}
      1\\
      0
    \end{bmatrix}\\
    \begin{bmatrix}
      y(k+1)\\
      z(k+1)
    \end{bmatrix}
    &=
    \begin{bmatrix}
      1 & 0 \\
      1 & 1
    \end{bmatrix}
    \begin{bmatrix}
      y(k)\\
      z(k)
    \end{bmatrix}
    +
    \begin{bmatrix}
      x(k)^2 + 1\\
      3w(k) + x(k)
    \end{bmatrix}
  \end{align*}
  This system has the closed-form solution
  \begin{align*}
    w(k) &= w(0) + \frac{(2^k - 1)}{3}x(0) + k \qquad x(k) = 2^kx(0) \\
    y(k) &= \frac{4^k-1}{3}x(0)^2 + y(0) + k\\
    z(k) &= 3w(0) + \frac{4^k-3k-1}{9}x(0)^2 + \\ 
    &\qquad (2^{k+1}-k-1)x(0) + ky(0) + z(0) + 2(k^2 - k)\ .
  \end{align*}
\end{example}



\section{Technical Details}
\label{Se:TechnicalDetails}

This section gives algorithms for summarizing recursive procedures using recurrence solving.
We assume that before these algorithms are applied to the procedures of a
program $\mathcal{P}$, we first compute and collapse the strongly connected
components of the call graph of $\mathcal{P}$ and topologically sort the 
collapsed graph.
Our analysis then works on the
strongly connected components of the call graph in a single pass,
in a topological order of the collapsed graph,
by applying the algorithms of this section to recursive components, and
applying intraprocedural analysis to non-recursive components.

For simplicity, \sectref{Height} focuses on the analysis of strongly connected
components consisting of a single recursive procedure $P$.
The first step of the analysis is to apply \algref{candidates}, which produces
a set of inequations that describe the values of variables in $P$.
Not all of the inequations found by \algref{candidates} are suitable for use in
a recurrence-based analysis, so we \OnlyPaper{apply a fixpoint
algorithm}\OnlyTech{apply \algref{filtering}} to filter the set of inequations
down to a subset that, when combined, form a stratified recurrence.  
The next step is to give this recurrence to a recurrence solver, which results
in a logical formula relating the values of variables in $P$ to the stack
height $h$ that may be used by $P$.  
In \sectref{DepthBound}, we show how to (i) obtain a bound on $h$ that depends
on the program state before the initial call to $P$, and (ii) combine the
recurrence solution with that depth bound to create a summary of $P$.
\OnlyTech{In \sectref{DualHeight}, we discuss how to obtain a certain class of more
precise bounds (including lower bounds on the running time of a procedure).}
In \sectref{Mutual}, we show how to extend the techniques of \sectref{Height} to
handle programs with mutual recursion, i.e., programs whose call graphs have
strongly connected components consisting of multiple procedures.
\OnlyTech{In \sectref{MissingBaseCases}, we discuss an extension of the
algorithm of \sectref{Mutual} that handles sets of mutually recursive
procedures in which some procedures do not have base cases.}
\OnlyPaper{(In the technical report version of this document, there are two
additional sub-sections.  
\cite[\S 4.3]{arxiv:BCKR2020} discusses a modified version of the algorithm of
\sectref{Height} that, in combination with the algorithm of \sectref{Height},
can prove more precise properties, e.g., that a variable is equal to, and not
only bounded by, some function of the depth of recursion.
\cite[\S 4.5]{arxiv:BCKR2020} discusses an extension of the algorithm of
\sectref{Mutual} that handles sets of mutually recursive procedures in which
some procedures do not have base cases.)}

\subsection{Height-Based Recurrence Analysis}
\label{Se:Height}

Let $\tau$ be a relational expression and let $P$ be a procedure.
We use $V_\tau(P,h)$ to denote the set of values of $\tau$ in a 
height-$h$ execution of $P$.
\[ V_\tau(P,h) \eqdef \{ \expsem{\tau}(\sigma,\sigma') : \tuple{\sigma,\sigma'} \in R(P,h) \} \]
It consists of values to which $\tau$ may evaluate at a state pair
belonging to $R(P,h)$.  
We call \textit{$\bsymGeneral{\tau} : \mathbb{N} \rightarrow \mathbb{Q}$ 
a bounding function for $\tau$ in $P$} if for all $h \in \mathbb{N}$
and all $v \in V_\tau(P,h)$, we have $v \leq
\bsym{\tau}{h}$.  Intuitively, the bounding function $\bsym{\tau}{h}$ 
bounds the value of an expression $\tau$ in any execution that uses
stack height at most $h$.  

The goal of \sectref{Height} is to find a set of relational
expressions and associated bounding functions. We proceed in three
steps.  First, we determine a set of candidate relational expressions
$\tau_1,...,\tau_n$.  Second, we optimistically assume that there
exist functions $\bsym{1}{h},...,\bsym{n}{h}$ that bound
these expressions, and we analyze $P$ under that assumption to obtain
constraints relating the values of the relational expressions to the values of the
$\bsym{1}{h},...,\bsym{n}{h}$ functions.  Third, we re-arrange
the constraints into recurrence relations for each of the
$\bsym{k}{h}$ functions (if possible) and solve them to
synthesize a closed-form expression for $\bsym{k}{h}$ that is
suitable to be used in a summary for $P$.

We begin our analysis of $P$ by determining a set of suitable expressions $\tau$.
If a relational expression $\tau$ has an associated bounding function, then it
must be the case that $V_\tau(P,1)$ (i.e., the set of values that $\tau$ takes
on in the base case) is bounded above.  Without loss of generality, we choose
expressions $\tau$ so that $V_\tau(P,1)$ is bounded above \textit{by zero}.  (Note
that if $V_\tau(P,1)$ is bounded above by $c$ then $V_{\tau-c}(P,1)$ is bounded
above by zero.)
We begin our analysis of $P$ by analyzing the base case to look for relational
expressions that have this property.

\begin{algorithm}[tb]
{\small
  \SetKwInOut{Input}{Input}
  \SetKwInOut{Output}{Output}
  \Input{A procedure $P$, and the associated vocabulary of program variables $\vars$}
  \Output{Height-based-recurrence summary $\summaryheight$}
  $\summarybase \gets \textit{Summary}(P, \textit{false})$ \label{Li:GetBase}\;
  $\wedgebase \gets \textit{Abstract}(\summarybase,\vars \cup \vars')$ \label{Li:GetWedgeBase}\;
  $n \gets$ the number of inequations in $\wedgebase$\;
  \ForEach{$k$ in $1,...,n$}{
      Let $\tau_k$ be the expression over $\vars \cup \vars'$ such that the $k^{\textrm{th}}$ inequation in 
      $\wedgebase$ is $(\tau_k \leq 0)$ \label{Li:CreateTau} \;
      Let $\bsym{k}{\cdot}$ be a fresh uninterpreted function symbol \label{Li:CreateSymbols}\;
  }
  $\summarycall \gets \bigwedge_{k=1}^{n} (\tau_k \leq \bsym{k}{h} \land \bsym{k}{h} \geq 0)$ \label{Li:SummaryCall}\;
  $\summaryrec \gets \textit{Summary}(P, \summarycall)$ \label{Li:SummaryRec}\;
  $\summaryextract 
    \gets \summaryrec \land \bigwedge_{k=1}^{n} (\bsym{k}{h+1} = \tau_k)$ \label{Li:SummaryExt}\;
  $S \gets \emptyset$\;
  \ForEach{$k$ in $1,...,n$}{
     $\wedgeextract{k} 
       \gets \textit{Abstract}(\summaryextract, \{ \bsym{1}{h},...,\bsym{n}{h},\bsym{k}{h+1} \})$
       \label{Li:WedgeExtract}\;
     \ForEach{inequation $\mathcal{I}$ in $\wedgeextract{k}$}{
           $S \gets S \cup \{\mathcal{I}\}$
     }
  } \label{Li:LastLine}
  \Return{$S$}
}
\caption{Algorithm for extracting candidate recurrence inequations \label{Alg:candidates}}
\end{algorithm}

\paragraph{Selecting candidate relational expressions.} 
The reason for
looking at expressions over program variables, as opposed to individual variables, 
is illustrated by \exref{SubsetSum}:
the variable $\prog{nTicks}$ has a different value each time the base case
executes, but the expression $\prog{nTicks}'-\prog{nTicks}-1$
is always equal to zero in the base case.

With the goal of identifying relational expressions that are bounded above by
zero, \algref{candidates} begins by
extracting a transition formula $\summarybase$ for the non-recursive paths
through $P$ by calling $\textit{Summary}(P,\textit{false})$ (i.e.,
summarizing $P$ by using $\textit{false}$ as a summary for the recursive
calls in $P$).
Next, we compute a set $\wedgebase$ of polynomial inequations over $\vars
\cup \vars'$ (the set of un-primed (pre-state) and primed (post-state) copies
of all global variables, along with unprimed copies of the parameters to $P$
and the variable $\prog{return}'$, which represents the return value of $P$)
that are implied by $\summarybase$ by calling 
$\textit{Abstract}(\summarybase, \vars \cup \vars')$.
\Omit{
Next, the formula $\summarybase$ is given as
input to $\textit{Abstract}$, on \lineref{GetWedgeBase}.
The second parameter to $\textit{Abstract}$ is the vocabulary
$\vars \cup \vars'$.

Calling $\textit{Abstract}$ on $\summarybase$ with the vocabulary
$\vars \cup \vars'$ produces a conjunction of inequations $\wedgebase$.
The inequations of $\wedgebase$ describe the state transformation performed by any
possible execution of the base case of $P$.}
Let $n$ be the number of 
inequations in $\wedgebase$.  Then, for $k=1,...,n$, we rewrite the $k^{th}$
inequation in the form $\tau_k \leq 0$.
In the case of \exref{SubsetSum}, 
$\tau_1 \eqdef \prog{return}'$ and 
$\tau_2 \eqdef \prog{nTicks}'-\prog{nTicks}-1$
have the property that $\tau_1 \leq 0$ and $\tau_2 \leq 0$ in the base case.

Note that there are, in general, many sets of relational expressions
$\tau_1,...,\tau_n$ that are bounded above by zero in the base case.  The
soundness of \algref{candidates} only depends on $\textit{Abstract}$ choosing
\textit{some} such set.  
Our implementation of $\textit{Abstract}$ uses
\cite[Alg.~3]{PACMPL:KCBR18}, and
is not guaranteed to choose the set of relational expressions that would lead
to the most precise results for any given application, e.g., for a given
assertion-checking or complexity-analysis problem.
Intuitively, in the case that $\summarybase$ is a
formula in linear arithmetic, our implementation of $\textit{Abstract}$
amounts to using the operations of the polyhedral abstract domain to find a
convex hull of $\summarybase$. 
Then, each of the inequations in the constraint representation of the convex
hull can be interpreted as a relational expression that is bounded above by
zero in the base case.

\paragraph{Generating constraints on bounding functions.}
For each of the expressions $\tau_k$ that has an upper bound in the base case, we
are ultimately looking to find a function $\bsym{k}{h}$ that is an upper bound
on the value of that expression in any height-$h$ execution.
Our way of finding such a function is to analyze the recursive cases of $P$ to
look for an invariant inequation that gives an upper bound on $V_{\tau_k}(P,h+1)$
in terms of an upper bound on $V_{\tau_k}(P,h)$.  Such an inequation can
be interpreted as a recurrence relating
$\bsym{k}{h+1}$ to $\bsym{k}{h}$.

The remainder of \algref{candidates} (\Lineseqref{SummaryCall}{LastLine}) finds
such invariant inequations.
The first step is to create the \textit{hypothetical procedure summary}
$\summarycall$, which hypothesizes that a bounding function $\bsymGeneral{\tau_k}$
exists for each expression $\tau_k$, and that the value of that function at
height $h$ is an upper bound on the value of $\tau_k$.
$\summarycall$ is a transition formula that represents a height-$h$
execution of $P$.
In \exref{SubsetSum}, $\summarycall$ is:
\begin{align*}
\prog{return}' &\leq \bsym{1}{h} 
\land 
\prog{nTicks}'-\prog{nTicks}-1 \leq \bsym{2}{h}
\land \\
\bsym{1}{h} &\geq 0
\land 
\bsym{2}{h} \geq 0
\end{align*}

On \lineref{SummaryRec}, \algref{candidates} calls $\textit{Summary}$, using
$\summarycall$ as the representation of
each recursive call in $P$, and the resulting transition formula is stored
in $\summaryrec$.  
Thus, $\summaryrec$ describes a typical height-$(h+1)$ execution of $P$.
In \exref{SubsetSum}, a simplified version of $\summaryrec$ is given as
$\varphi_{h+1}$ in \sectref{Overview}.

On \lineref{SummaryExt}, the formula $\summaryextract$ is produced by 
conjoining $\summaryrec$ with a formula stating that, for each $k$,
$\bsym{k}{h+1}=\tau_k$.
Therefore, $\summaryextract$ implies that any upper bound on $\bsym{k}{h+1}$
must be an upper bound on $\tau_k$ in any height-$(h+1)$ execution.  

Ultimately, we wish to obtain a closed-form solution for each $\bsym{k}{h}$.
The formula $\summaryextract$ \emph{implicitly} determines a set of recurrences
relating $\bsym{1}{h+1},...,\bsym{n}{h+1}$ to $\bsym{1}{h},...,\bsym{n}{h}$.
However, $\summaryextract$ does not have the \emph{explicit} form of a 
recurrence.  
\Lineseqref{WedgeExtract}{LastLine} abstract $\summaryextract$ to a
conjunction of inequations 
that give an explicit relationship between $\bsym{k}{h+1}$ and 
$\bsym{1}{h},...,\bsym{n}{h}$ for each $k$.

\OnlyTech{
\begin{algorithm}[h]
{\small
  \SetKw{Continue}{continue}
  \SetKwInOut{Input}{Input}
  \SetKwInOut{Output}{Output}
  \Input{A set of candidate inequations $\mathcal{I}_1,...,\mathcal{I}_N$
         over the function symbols $\bsym{1}{h},...,\bsym{n}{h},\bsym{1}{h+1},...,\bsym{n}{h+1}$}
  \Output{A set of inequations that form a stratified recurrence}
  Let $\boundsmap[j]$ be a map from integers to integers\;
  Let $\depsmap[j,k]$ and $\nonlindepsmap[j,k]$ be maps that map all pairs of integers 
    to \textit{false}\;
  $S \gets \{1,...,N\}$ \;
  \ForEach{$j$ in $1,...,N$}{
    Write $\mathcal{I}_j$ as $\bsym{k}{h+1} \leq c_0 + 
      \Sigma_{i=1}^{m_j} c_i (\bsym{1}{h})^{d_{i,1}} \cdots (\bsym{n}{h})^{d_{i,n}}$
      if $\mathcal{I}_j$ can be written in that form with
      $1 \leq k \leq n$, $\forall i.~c_i \in \mathbb{Q}$, $\exists i > 0.~c_i > 0$,
      $\forall i,p.~d_{i,p} \in \mathbb{N}$; 
      otherwise let $S \gets S - \{j\}$ and \Continue loop\;
    For $i=0,...,m_j$, $c_i' \gets \max(0,c_i)$ \label{Li:DropNegative}\;
    Let $\mathcal{I}_j'$ be
      $\bsym{k}{h+1} \leq c_0' + 
      \Sigma_{i=1}^{m_j} c_i' (\bsym{1}{h})^{d_{i,1}} \cdots (\bsym{n}{h})^{d_{i,n}}$\;
    $\boundsmap[j] := k$\;
    \ForEach{$i \in \{1,..,m_j\}$}{
      \ForEach{$p \in \{1,...,n\}$}{
        \lIf{$c_i' > 0 \land d_{i,p} > 0$}{
          $\depsmap[j,p]:=\textit{true}$
        }
        \lIf{$c_i' > 0 \land d_{i,p} > 0 \land \Sigma_{q=1}^n d_{i,q} > 1$}{
          $\nonlindepsmap[j,p]:=\textit{true}$
        }
      }
    }
  }
  $A \gets \emptyset$ \;
  \Repeat{$V = \emptyset$}{
    $V \gets S - A$\;
    \Repeat{$V$ is unchanged}{
      \ForEach{$j \in V$}{
         \lIf{$\exists k. \depsmap[j,k] \land
                            \lnot \exists j' \in V. \boundsmap[j']=k$ 
                            \label{Li:TransitiveCriterion}}{$V \gets V - \{j\}$}
         \lIf{$\exists k. \nonlindepsmap[j,k] \land
                            \lnot \exists j' \in A. \boundsmap[j']=k$ 
                            \label{Li:NonLinearCriterion}}{$V \gets V - \{j\}$}
      }
    }
    \ForEach{$k=\{1,...,n\}$}{
      \If{$V$ contains more than one $j$ such that $\boundsmap[j]=k$}{
        Arbitrarily choose one such $j$ to remain in $V$, and remove all other such $j$ from $V$
        \label{Li:ArbitraryChoice}\;}
    }
    $A \gets A \cup V$\;
  }
  \Return{$\{\mathcal{I}_j' \mid j \in A\}$}
}
\caption{Algorithm for constructing a stratified recurrence \label{Alg:filtering}}
\end{algorithm}
}

\paragraph{Extracting and solving recurrences.}  The next step of height-based
recurrence analysis is to identify a subset of the
inequations returned by \algref{candidates} that constitute a 
stratified system of polynomial recurrences (\defref{solvable}).  
This subset must meet the following three \textit{stratification criteria}:

\begin{enumerate}
\item Each bounding function $\bsym{k}{h+1}$ must appear on the
  left-hand-side of at most one inequation.
\item If a bounding function $\bsym{k}{h}$ appears on the
  right-hand-side of an inequation, then $\bsym{k}{h+1}$ appears on
  some left-hand-side.
\item It must be possible to organize the $\bsym{k}{h+1}$ into
  \textit{strata}, so that if $\bsym{k}{h}$ appears in a non-linear
  term on the right-hand-side of the inequation for $\bsym{j}{h+1}$, then
  $\bsym{k}{h}$ must be on a strictly lower stratum than $\bsym{k}{h}$.
\end{enumerate}

\OnlyPaper{A maximal subset of inequations that complies with the above three
rules can be computed in polytime using a fixpoint algorithm.  (An algorithm
for extracting a stratified recurrence is given in the technical-report version
of this document as \cite[Alg.~3]{arxiv:BCKR2020}.)}
\OnlyTech{\algref{filtering} computes a \textit{maximal} subset of inequations
that complies with the above three rules.}

The next step of height-based
recurrence analysis is to send this recurrence to a recurrence solver, such
as the one described in \citet{PACMPL:KCBR18}.  
The solution to the recurrence is a set of bounding functions. 
Let $B$ be the set of indices $k$ such that we found a recurrence for,
and obtained a closed-form solution to, the bounding function $\bsym{k}{h}$.
Using these bounding functions, we can derive the following procedure summary
for $P$, which leaves the height $H$ unconstrained.
\begin{align}
  \label{Eq:NoHeightSummary}
  \exists H .
  \bigwedge_{k \in B}
  [
  \tau_k \leq \bsym{k}{H}
  ]
\end{align}
The subject of \sectref{DepthBound} is to find
a formula 
$\summarydepth{P_i}(H,\sigma)$
relating $H$ to the pre-state $\sigma$ of the initial call to $P$.
The formula 
$\summarydepth{P_i}(H,\sigma)$
can be combined
with \eqref{NoHeightSummary} to obtain a more precise procedure summary.

\newcommand{\iffdef}{\buildrel \mbox{\tiny\rm def} \over \iff}
\paragraph{Soundness.} 
Roughly, the soundness of height-based recurrence analysis follows from: (i)
sound extraction of the recurrence constraints used by CHORA to characterize
non-linear recursion; (ii) sound recurrence solving; and (iii) soundness of the
underlying framework of algebraic program analysis.
The soundness of parts (ii) and (iii) depends on the soundness of prior work
\cite{PACMPL:KCBR18}.
\OnlyPaper{The soundness of (i) is addressed in a detailed proof in
the appendix of the technical report version of this document
\cite{arxiv:BCKR2020}.}
\OnlyTech{The soundness of (i) is addressed in a detailed proof in
the appendix of this document.}
The soundness property proved there is as follows:
let $P$ be a procedure to which \algref{candidates} and 
\OnlyTech{\algref{filtering}}\OnlyPaper{the recurrence-extraction algorithm} have
been applied to obtain a stratified recurrence.
Let $\{\tau_i\}_{i \in [1,n]}$ be the relational expressions computed by
\algref{candidates}.
Let $B \subseteq [1,n]$ be such that $\{b_i\}_{i \in B}$ is 
the set of functions produced by solving the stratified recurrence.
We show that each $b_i$ function bounds the corresponding $V_{\tau_i}(P,h)$
value set.
In other words, the following statement holds:
$\forall h \geq 1 . \bigwedge_{i \in B} \forall v \in V_{\tau_i}(P,h). v \leq b_i(h)$.

\subsection{Depth-Bound Analysis}
\label{Se:DepthBound}

\begin{algorithm}[t]
{\small
  \SetKwInOut{Input}{Input}
  \SetKwInOut{Output}{Output}
  \Input{A weighted control-flow graph $(V,\weightededges,\calledges)$}
  \Output{Depth-bound formulas $\summarydepth{P_1}(D,\sigma),...,\summarydepth{P_n}(D,\sigma)$}
  \ForEach{$i\in \{1,...n\}$}{Let $\sccentry{P_i}$ be a new vertex}\label{Li:DepthFirstLine}
  Let $\sccexit$ be a new vertex\;
  $V' \gets V \cup \{\sccexit\} \cup \{\sccentry{P_i} \mid i \in \{1,...,n\}\}$\;
  Create a new integer-valued auxiliary variable $D$\;
  $\newedges \gets \weightededges$\;
  \ForEach{$i \in \{1,...,n\}$}{
    $\newedges \gets \newedges \cup \{(\sccentry{P_i}, \depthone, \ventry{P_i})$\}
    $\cup \{(\ventry{P_i}, \summarybasefor{P_i}, \sccexit)$\}
  }
  \ForEach{call edge $(u,Q,v)$ in $\calledges$}{
    \If{$Q=P_i$ for some $i$}{
      $\newedges \gets \newedges \cup \{(u,\depthinc,\ventry{Q})\}
      \cup \{(u,\varphi_{[havoc]},v)\}$
    }
    \Else{
      $\newedges \gets \newedges \cup \{(u,\summaryfor{Q},v)\}$ \label{Li:DepthBeforeSummary}
    }
  }
  \ForEach{$i=1,...,n$}{
    $\summarydepth{P_i}(D,\sigma) 
         \gets \textrm{PathSummary}\,(\sccentry{P_1},\sccexit,
                                      V',\newedges,\emptyset)$
                                      \label{Li:PathSummary}
  }
  \Return{$\summarydepth{P_1}(D,\sigma),...,\summarydepth{P_n}(D,\sigma)$}
}
\caption{Algorithm for producing a depth-bound formula \label{Alg:depth}}
\end{algorithm}

In \sectref{Height}, we showed how to find a bounding function
$\bsym{\tau}{h}$ that gives an upper bound on the value of a relational
expression $\tau$ in an execution of a procedure $P_i$ as a function of the
stack height (i.e., maximum depth of recursion) $h$ of that execution.
In this section, the goal is to find bounds on the maximum depth of
recursion $h$ that may occur as a function of the pre-state $\sigma$
(which includes the values of global variables and parameters to $P_i$) from
which $P_i$ is called.

For example, consider \exref{SubsetSum}.  The algorithms of \sectref{Height}
determine bounds on the values of two relational expressions in terms of $h$,
namely:
$\prog{nTicks}' \leq \prog{nTicks} + 2^{h} - 1$, and 
$\prog{return}' \leq h - 1$.
The algorithm of this sub-section (\algref{depth}) determines that
$h$ satisfies
$h \leq \max(1,1+\prog{n}-\prog{i})$.
These facts can be combined to form a procedure summary for
\textit{SubsetSumAux} that relates the return value and the increase to
$\prog{nTicks}$ to the values of the parameters $\prog{i}$ and $\prog{n}$.

The stack height $h$ required to execute a procedure often depends on the
number of times that some transformation can be applied to the procedure's
parameters before a base case must execute.
For example, in \exref{SubsetSum}, the height bound is a consequence of the
fact that $\prog{i}$ is incremented by one at each recursive call, until
$\prog{i} \geq \prog{n}$, at which point a base case executes.
Likewise, in a typical divide-and-conquer algorithm, a size parameter is
repeatedly divided by some constant until the size parameter is below some
threshold, at which point a base case executes.  
Intuitively, the technique described in this section is designed to discover
height bounds that are consequences of such repeated transformations (e.g.,
addition or division) applied to the procedures' parameters.

To achieve this goal, we use \algref{depth}, which is inspired by the algorithm
for computing bounds on the depth of recursion in \citet{TCL:AGM13}.
\algref{depth} constructs and analyzes an over-approximate \emph{depth-bounding
model} of the procedures $P_1,...,P_n$ that includes an auxiliary depth-counter
variable, $D$. 
Each time that the model descends to a greater depth of recursion, $D$ is
incremented.
The model exits only when a procedure executes its base case.  
In any execution of the model, the final value of $D$ thus represents the depth
of recursion at which some procedure's base case is executed.  

\algref{depth} takes as input a representation of the procedures in $S$ as a
single, combined control-flow graph $(V,\weightededges,\calledges)$ having two
kinds of edges: (1) weighted edges $(u,\varphi,v) \in \weightededges$, which
are weighted with a transition formula $\varphi$, and (2) call edges in the set
$\calledges$.  Each call edge in $\calledges$ is a triple $(u,Q,v)$, in which
$u$ is the call-site vertex, $v$ is the return-site vertex, and the edge is
labeled with $Q$, representing a call to a procedure $Q$.
We assume that if any procedure $Q \notin S$ is called by some procedure in
$S$, then $Q$ has been fully analyzed already, and therefore a procedure
summary $\summaryfor{Q}$ for $Q$ has already been computed.
Each procedure $Q$ has an entry vertex $e_Q$, an exit vertex $x_Q$, and
a transition formula $\beta_Q$ that over-approximates the base cases of $Q$.
Note that $(V,\weightededges,\calledges)$ consists of several disjoint,
single-procedure control-flow graphs when $n>1$.

On \lineseqref{DepthFirstLine}{DepthBeforeSummary}, \algref{depth} constructs
the depth-bounding model, represented as a new control-flow graph
$(V',\newedges,\emptyset)$. 
The algorithm begins by creating new auxiliary entry vertices 
$\sccentry{P_1},...,\sccentry{P_n}$ for the procedures $P_1,...,P_n$ and
a new auxiliary exit vertex $\sccexit$.  The new vertex set $V'$ contains
$V$ along with these $n+1$ new vertices.  \algref{depth} then creates a
new integer-valued variable $D$.
For $i=1,...,n$, the algorithm then creates an edge from $\sccentry{P_i}$ to
$\ventry{P_i}$, weighted with a transition formula that initializes
$D$ to one, and an edge from $\vexit{P_i}$ weighted with the formula
$\summarybasefor{P_i}$, which is a summary of the base case of $P_i$.  

\algref{depth} replaces every call edge $(u,Q,v) \in \calledges$ with one or
more weighted edges.  Each call to a procedure $Q \notin \{P_1,...,P_n\}$
is replaced by an edge $(u,\summaryfor{Q},v)$ weighted with the procedure
summary $\summaryfor{Q}$ for $Q$.  Each call to some $P_i$ is replaced by
two edges.  The first edge represents descending into $P_i$, and goes from $u$
to $\ventry{P_i}$, and is weighted with a formula that increments $D$
and havocs local variables.  
The second edge represents skipping over the call to $P_i$ rather than
descending into $P_i$.  This edge is weighted with a transition formula
that havocs all global variables and the variable $\prog{return}$, but leaves
local variables unchanged.  

The final step of \algref{depth}, on \lineref{PathSummary}, actually computes
the depth-bounding summary $\summarydepth{P_i}(D,\sigma)$ for each procedure $P_i$.
Because there are no call edges in the new control-flow graph
$(V',\newedges,\emptyset)$, intraprocedural-analysis techniques can be used to
compute transition formulas that summarize the transition relation for all paths
between two specified vertices.  For each procedure $P_i$, the formula
$\summarydepth{P_i}(D,\sigma)$ is a summary of all paths from $\sccentry{P_i}$ to
$\sccexit$, which serves to relate $D$ to $\sigma$, which is the pre-state of
the initial call to $P_i$.

The formulas $\summarydepth{P_i}(D,\sigma)$ for $i=1,...,n$ can be used to
establish an upper bound on the depth of recursion in the following way.
Let $(\sigma,\sigma')$ be a state pair in the relational semantics
$\relsem{P_i}$ of $P_i$.
Then, there is an execution $e$ of $P_i$ that starts in state $\sigma$ and finishes
in state $\sigma'$, in which the maximum\footnote{ Note that non-terminating
executions of $P_i$ do not correspond to any state-pair $(\sigma,\sigma')$ in
the relational semantics $\relsem{P}$; therefore, such executions are not
represented in the procedure summary for $P_i$ that we wish to construct.}
recursion depth is some $d \in \mathbb{N}$.
Then there is a path through the control-flow graph $(V',\newedges,\emptyset)$
that corresponds to the path taken in $e$ to reach some execution of a base
case at the maximum recursion depth $d$.  
Therefore, if $d$ is a possible depth of recursion when starting from state
$\sigma$, then there is a satisfying assignment of
$\summarydepth{P_i}(D,\sigma)$ in which $D$ takes the value $d$.
The contrapositive of this argument says that, if there does not exist any
satisfying assignment of $\summarydepth{P_i}(D,\sigma)$ in which $D$ takes
the value $d$, then it must be the case that no execution of $P_i$ that
starts in state $\sigma$ can have maximum recursion depth $d$.  In this way,
$\summarydepth{P_i}(D,\sigma)$ can be interpreted as providing bounds on
the maximum recursion depth that can occur when $P_i$ is started in state
$\sigma$.

Once we have the depth-bound summary $\summarydepth{P}$ for some procedure
$P$, we can combine it with the closed-form solutions for bounding functions
that we obtained using the algorithms of \sectref{Height} to produce a
procedure summary.  Let $B$ be the set of indices $k$ such that we found a
recurrence for the bounding function $\bsym{k}{h}$.  We produce a
procedure summary of the form shown in \eqref{HBASummary}, which uses the
depth-bound summary $\summarydepth{P}$ to relate the pre-state $\sigma$ to
the variable $H$, which in turn is used to index into the bounding function
$\bsym{k}{h}$ for each $k \in B$.
\begin{align}
  \label{Eq:HBASummary}
  \exists H .
  \summarydepth{P}(H,\sigma)
  \land
  \bigwedge_{k \in B}
  [
  \tau_k \leq \bsym{k}{H}
  ]
\end{align}

\OnlyTech{
\subsection{Finding Lower Bounds Using Two-Region Analysis}
\label{Se:DualHeight}

In this sub-section, we describe an extension of height-based recurrence analysis, called \textit{two-region analysis}, that is able to prove stronger
conclusions, such as non-trivial lower bounds on the running times of some
procedures.  

In \sectref{Height}, we discussed height-based recurrence analysis, and showed
how it can find an upper bound on the increase to the variable $\prog{nTicks}$ in
\exref{SubsetSum}.  
Now, we consider the application of height-based recurrence analysis to the
procedure \textit{differ} shown in \figref{Differ}.  \textit{differ} uses the
global variables $\prog{x}$ and $\prog{y}$ to (in effect) return a pair of 
integers.
The pair $(\prog{x},\prog{y})$ returned by the procedure is formed from the
$\prog{x}$ value returned by the first call and the $\prog{y}$ value returned
by the second call, each incremented by one.
The base case occurs when the parameter $\prog{n}$ equals zero or one, and at
each call site, the parameter $\prog{n}$ is decreased by either one or two.
We will apply height-based recurrence analysis and two-region analysis to
look for bounds on $\prog{x}'$ and $\prog{y}'$, and their sum and difference,
after \textit{differ} is called with a given value
$\prog{n}$.

For the purposes of the following discussion, we will focus on $\prog{x}$,
but the same conclusions apply to $\prog{y}$.
By applying height-based recurrence analysis to the procedure \textit{differ},
we can prove that the post-state value $\prog{x}'$ is upper-bounded by
$\prog{n}-1$.
At the same time, the analysis also proves a \textit{lower} bound on
$\prog{x}'$ by considering the term $\tau_1=-\prog{x}'$.  However, the bounding
function $\bsym{1}{h}$ obtained by height-based analysis is the constant
function $\bsym{1}{h} = 0$, which yields the trivial lower bound 
$\prog{x}' \geq 0$.  As a result, the results of height-based recurrence
analysis can only be used to prove that the difference between $\prog{x}'$
and $\prog{y}'$ is at most $\prog{n}-1$, which is an over-estimate by a 
factor of two.

\begin{figure}[t]
  $\begin{array}{@{\hspace{0ex}}c@{\hspace{5.0ex}}c@{\hspace{0ex}}}
  {
    \begin{array}{l}
    \hspace{0.0cm} \textrm{int}~x;~\textrm{int}~y;                                      \\
    \hspace{0.0cm} \textrm{int}~\textit{differ}(\textrm{int}~n)~\{                      \\
    \hspace{0.5cm}     \textbf{if}~(n == 0~||~n == 1)~\{~ x=0;~y=0;~\textbf{return}; \} \\
    \hspace{0.5cm}     \textit{differ}(nondet()~?~n - 1 : n - 2);                       \\ 
    \hspace{0.5cm}     \textrm{int}~temp~=~x; \textit{// Store x ``returned'' by first call} \\
    \hspace{0.5cm}     \textit{differ}(nondet()~?~n - 1 : n - 2);                       \\
    \hspace{0.5cm}     x = temp + 1; y = y + 1; \textit{// ``Return'' (temp+1,y+1)}     \\
    \hspace{0.0cm} \}                                                           
    \end{array}
  } \\
  { \parbox[c]{5cm}{ \includegraphics[width=2in]{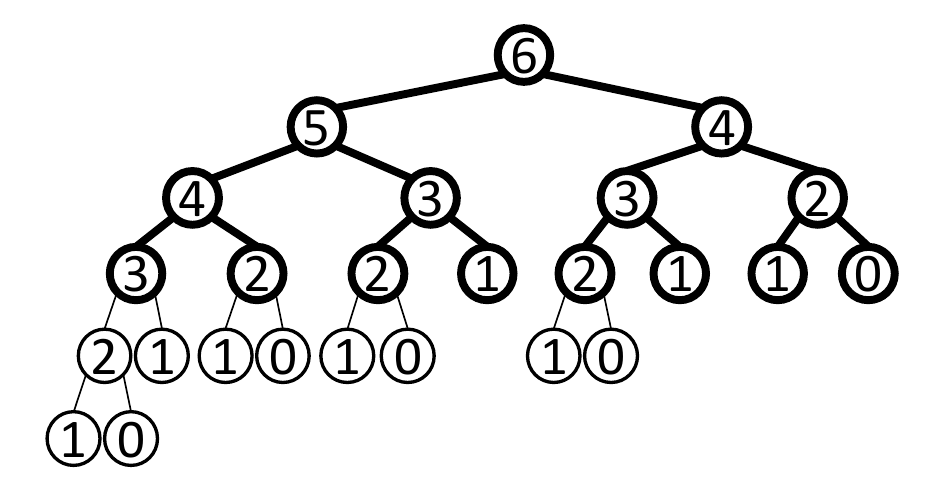} } }
  \end{array}$
\caption{\label{Fi:Differ}
Source code for the non-linearly recursive procedure \textit{differ},
and a depiction of a possible tree of recursive calls \textit{differ}.
The number inside each vertex indicates the value of the parameter $n$ at that
execution of \textit{differ}.  The minimum depth $M$ at which a base case
occurs in this recursion tree is 3.  The \textit{upper region} of the tree,
which includes all vertices at depth less than or equal to $M$, is shown with
bold outlines.
}
\end{figure}

In this sub-section, we extend our formal characterization of the relational
semantics of a procedure (given in \sectref{Background}) in the following way.
We use $V_\tau(R)$ to denote the set of values that $\tau$ takes on in a state
relation $R$.
That is, $V_\tau(R) \eqdef \{ \expsem{\tau}(\sigma,\sigma') : \tuple{\sigma,\sigma'} \in R \}$.
We view a procedure $P$ as a pair consisting of a state relation $\basesem{P}
\in 2^{\State{} \times \State{}}$ (which gives the relational
semantics of the ``base case'' of $P$) and a ($\bottom$-strict)
function $\relfunsemrec{P} : 2^{\State{} \times \State{}}
\rightarrow 2^{\State{} \times \State{}}$ (which gives the
relational semantics of the ``recursive case'' of $P$), such that
$\relfunsem{P}(X) = \relfunsemrec{P}(X) \cup \basesem{P}$ for any
state relation $X$.  
For any natural number $m$, define
$\relfunsemrec{P}^m$ to be the $m$-fold composition of
$\relfunsemrec{P}$, and define
$\relfunsem{P}^m$ to be the $m$-fold composition of
$\relfunsem{P}$.
Note that $\relfunsemrec{P}^m(\basesem{P})$ corresponds to the state
relation that is exactly $m$ ``steps'' away from $\basesem{P}$, whereas
$\relfunsem{P}^m(\basesem{P})$ corresponds to a state relation
that is inclusive of all state relations between zero and $m$ 
steps away from $\basesem{P}$.
We say that a function
\textit{$\bsymGeneral{\tau} : \mathbb{N} \times \mathbb{N}
  \rightarrow \mathbb{Q}$ is a lower bound for $\tau$ in $P$} if for
all $n,m \in \mathbb{N}$ and all $v \in
V_{\tau}(\relfunsemrec{P}^m(\relsem{P}(n)))$, we have 
$\bsym{\tau}{m,n} \leq v$.  Our goal in this sub-section is to find
such lower-bounding functions.

The preceding formal presentation can also be understood using the following
intuitive characterization of a tree of recursive calls.
We characterize a tree of recursive calls with two parameters:
(i) the height $H$, and (ii) the minimum depth $M$ at which a base case occurs.
Importantly, the depth-bound analysis of \sectref{DepthBound} can be
used to obtain bounds on the parameters $H$ and $M$.
We define the \textit{upper region} of $T$ to be the tree that is
produced by removing from $T$ all vertices that are at depth greater than
$M$.  The \textit{lower region} of $T$ contains all the vertices of $T$
that are not present in the upper region.  In general, the lower region
is comprised of zero or more trees.  (In \figref{Differ}, the upper
region is shown with bold outlines.)

The relational semantics of the lower region are given by 
$\relsem{P}(H-M)$, where $H-M$ is the maximum height of any vertex at
depth $M$, which corresponds to the bottom of the upper region.
The relational semantics of the upper region are given by 
$\relfunsemrec{P}^M(\relsem{P}(H-M))$.
The idea of our approach is to apply height-based recurrence analysis in the
lower region (to summarize $\relsem{P}(H-M)$), and a modified analysis in the
upper region (to summarize $\relfunsemrec{P}^M(X)$), and then combine the
results to produce a procedure summary
for $\relfunsemrec{P}^M(\relsem{P}(H-M))$.

In the lower region, we perform height-based recurrence analysis unmodified
(as in \sectref{Height}) to obtain, for each relational expression
$\tau_k$, a bounding function $\bsymsuper{k}{h}{L}$.
The only difference is that we will not evaluate our bounding functions
at the height $H$ of the entire tree to find bounds on the value of
$\tau_k$ at the root of the tree.  Instead, we use the lower-region
bounding functions $\bsymsuper{k}{h}{L}$ to obtain bounds on the value of
$\tau_k$ at the height $(H-M)$.

In the upper region, we perform a modified height-based recurrence
analysis in which we substitute the notion of \textit{upper-region height}
for the notion of height.  The upper-region height of a vertex $v$
at depth $d_v$ in the upper region is defined to be $M-d_v$.  Thus, vertices
at depth $M$ (i.e., the bottom of the upper region) have upper-region height
zero, and the root (at depth 0) has upper-region height $M$.
For each $\tau_k$, the upper-region bounding function $\bsymsuper{k}{h}{U}$
needs to bound $V_{\tau_k}(\relfunsem{P}^M(X))$.
Therefore, in the upper region, we only require the bounding function
$\bsymsuper{k}{h}{U}$ to be a bound on the values that the expression $\tau_k$
can take on at exactly the upper-region height $h$, rather than requiring
$\bsymsuper{k}{h}{U}$ to be an upper bound on the values that $\tau_k$ can 
take on at any height between one and $h$.
Consequently, bounding functions $\bsymsuper{k}{h}{U}$ are not required to be
non-decreasing as upper-region height increases.

We make three changes to the algorithms of \sectref{Height} to find the 
bounding functions $\bsymsuper{k}{h}{U}$ for the upper region.  
First, in \algref{candidates}, on \lineref{SummaryCall}, we remove the conjunct
that asserts that the bounding functions are greater than or equal to zero.
Second, in \algref{candidates}, we modify
\lineref{SummaryRec} so that the resulting summary formula $\summaryrec$ is a
summary of only the recursive paths through the
procedure\footnote{We obtain a summary that excludes non-recursive paths by
adding an auxiliary flag variable $r$ to the program that indicates whether a
recursive call has occurred, and then modifying our internal representation of
the procedure so that (i) $r$ is initially false, (ii) $r$ is updated to true
when a call occurs, and (iii) $r$ is assumed to be true at the end of the
procedure.}, 
rather than a summary that includes base cases.  Third, we change
\algref{filtering} by removing \lineref{DropNegative}, so that recurrences are
allowed to have a negative constant coefficient.  

Analysis results for the two regions are combined in the following way.  
After analyzing both regions, we have obtained, for several quantities 
$\tau_k$, closed-form solutions to the recurrences for two bounding functions.
$\bsymsuper{k}{h}{L}$ is the closed form solution for the
lower-region bounding function in terms of the height $h$.
The upper-region closed-form solution 
$\bsymsuper{k}{h',c_k^{U}}{U}$ 
is expressed in terms of two parameters:
an upper-region-height parameter $h'$, and a symbolic initial condition parameter
$c_k^U$ that determines the value of the bounding function when the 
upper-region-height parameter is zero.

We relate the values of the two bounding functions to one another and to
the associated term $\tau_k$ over program variables by constructing the
formula given below as \eqref{LinkedBounds}.  In \eqref{LinkedBounds},
bounding functions obtained by height-based analysis of the lower region
always equal zero at height one, just as in \sectref{Height}.  
By contrast, the initial condition parameter $c_k^{U}$ for the upper region is
specified to be $\bsymsuper{k}{H-M}{L}$, i.e.,  the value of the lower-region
bounding function evaluated at height $H-M$.

As in \sectref{DepthBound}, we use, for each procedure $P$, the depth-bound
formula $\summarydepth{P}(D,\sigma)$ to bound the tree-shape parameters $H$ and
$M$ as a function of the pre-state $\sigma$ of the initial call to $P$.  In effect, 
$\summarydepth{P}(D,\sigma)$ constrains its parameter $D$ to equal the length
of some feasible root-to-leaf path in a tree of recursive calls starting from
$\sigma$.
Thus, we can obtain a sound upper bound on $H$ and a sound lower bound on
$M$ by using two copies of $\summarydepth{P}(D,\sigma)$ instantiated with the
two shape parameters, because $H$ is upper-bounded by the length of the longest
root-to-leaf path in the tree of recursive calls, and $M$ is lower-bounded by
the length of the shortest root-to-leaf path.  

As in the earlier procedure summary formula \eqref{HBASummary} in
\sectref{DepthBound}, $B$ represents the set of indices $k$ such that we
obtained bounding functions in both the lower and upper regions. 
The final procedure summary produced by two-region analysis is given
below as \eqref{LinkedBounds}.
\begin{multline}
  \label{Eq:LinkedBounds}
  \exists H . \exists M .
  M \leq H \land
  \summarydepth{P}(M,\sigma)
  \land
  \summarydepth{P}(H,\sigma)
  \land \\
  \bigwedge_{k \in B}
  [
  \tau_k \leq \bsymsuper{k}{M,\bsymsuper{k}{H-M}{L}}{U}
  ]
\end{multline}

We now consider the application of \eqref{LinkedBounds} to the procedure
\textit{differ} from \figref{Differ}.  
The two bounded terms related to $\prog{x}'$ are $\tau_1=-\prog{x}'$ and
$\tau_2=\prog{x}'$.  
(There are also two terms for $\prog{y}'$ that are analogous to those for
$\prog{x}'$.)
The lower-region and upper-region recurrences for
these terms are as follows.

\noindent\begin{minipage}{.5\linewidth}
\begin{align}
\bsymsuper{1}{h+1}{L} &= \bsymsuper{1}{h}{L}     \label{Eq:LowerRec} \\
\bsymsuper{1}{h'+1}{U} &= \bsymsuper{1}{h'}{U} - 1 \label{Eq:UpperRec}
\end{align}
\end{minipage}%
\begin{minipage}{.5\linewidth}
\begin{align}
\bsymsuper{2}{h+1}{L} &= \bsymsuper{2}{h}{L} + 1 \\ 
\bsymsuper{2}{h'+1}{U} &= \bsymsuper{2}{h'}{U} + 1
\end{align}
\end{minipage}

\smallskip
The closed-form solutions to these recurrences
are as follows.

\noindent\begin{minipage}{.5\linewidth}
\begin{align}
\bsymsuper{1}{h}{L} &= 0     \label{Eq:LowerSoln} \\
\bsymsuper{1}{h',c_1^U}{U} &= c_1^U - h' \label{Eq:UpperSoln}  
\end{align}
\end{minipage}%
\begin{minipage}{.5\linewidth}
\begin{align}
\bsymsuper{2}{h}{L} &= h \label{Eq:LowerUpper} \\
\bsymsuper{2}{h',c_2^U}{U} &= c_2^U + h' \label{Eq:UpperUpper}
\end{align}
\end{minipage}

\smallskip

A much-simplified version of the procedure summary that we obtain for \textit{Differ} is:
\begin{align}
  \label{Eq:DifferSummary}
  \frac{\prog{n}-1}{2} \leq \prog{x}' \leq \prog{n} 
  \land
  \frac{\prog{n}-1}{2} \leq \prog{y}' \leq \prog{n} 
\end{align}

The key difference between the upper and lower regions is that \eqref{LowerRec}
leads to the non-decreasing solution \eqref{LowerSoln}, whereas \eqref{UpperRec}
leads to the strictly decreasing solution \eqref{UpperSoln}.  
In the final procedure summary, the initial conditions in the lower region  
are be specified to equal zero.
Nevertheless, the lower-region recurrence solutions can create a non-zero gap
between the lower bound (\eqref{LowerSoln}) on $\prog{x}'$ and the upper bound
(\eqref{LowerUpper}) on $\prog{x}'$ (when $h > 0$).  In the upper region, the
solutions \eqref{UpperSoln} and \eqref{UpperUpper} represent a lock-step
increase in the upper and lower bounds on $\prog{x}'$ as $h'$ increases
(because $-c_1^U+h' \leq \prog{x}' \leq c_2^U+h'$ for any $h'$).
However, there can be a gap between the initial condition values 
$c_1^U=\bsymsuper{1}{h}{L} = 0$ and
$c_2^U=\bsymsuper{2}{h}{L} = h$.}

\subsection{Mutual Recursion}
\label{Se:Mutual}
In this section, we describe the generalization of the height-based recurrence
analysis of \sectref{Height} to the case of mutual recursion.  
Instead of analyzing a single procedure $P$, we assume that we are given a set
of procedures $P_1,...,P_m$ that form a strongly connected component of the
call graph of some program.

\begin{example} 
We use the following program to illustrate the application of our technique to 
mutually recursive procedures.  The procedure \textit{P1} increments the global
variable $\prog{g}$ in its base case, and calls \textit{P2} eighteen times
in a for-loop in its recursive case.  Similarly, \textit{P2} increments
$\prog{g}$ in its base case and calls \textit{P1} two times in a for-loop
in its recursive case.
\label{Exa:P1P2}
  $\begin{array}{l}
    \hspace{0.0cm} \textrm{int}~g;                                               \\
    \hspace{0.0cm} \textrm{void}~P1(\textrm{int}~n)~\{                              \\
    \hspace{0.5cm}     \textbf{if}~(n <= 1)~\{~ g\textrm{++}; \textbf{return};\} \\
    \hspace{0.5cm}     \textbf{for}(\textrm{int}~i=0;i<18;i\textrm{++})\{~P2(n-1);\}\\
    \hspace{0.0cm} \}                                                               \\
  \end{array}$
  \\
  $\begin{array}{l}
    \hspace{0.0cm} \textrm{void}~P2(\textrm{int}~n)~\{                              \\
    \hspace{0.5cm}     \textbf{if}~(n <= 1)~\{~ g\textrm{++}; \textbf{return};\} \\
    \hspace{0.5cm}     \textbf{for}(\textrm{int}~i=0;i<2;i\textrm{++})\{~P1(n-1);\} \\
    \hspace{0.0cm} \}                                                               \\
  \end{array}$
\end{example}

To apply height-based recurrence analysis to a set $S=\{P_1,...,P_m\}$ of mutually
recursive procedures, we use a variant of \algref{candidates} that interleaves
some of the analysis operations on the procedures in $S$.
Specifically, we make the following changes to \algref{candidates}.
First, we perform the operations on \lineseqref{GetBase}{SummaryCall} for each
procedure $P_i$ to obtain the symbolic summary formula $\summarycallfor{P_i}$.
For each procedure $P_i$, we obtain a set of bounded terms
$\tau_{i,1},...,\tau_{i,n_i}$, and our goal will be to find a height-based
recurrence for each such term.  

Note that a term $\tau_{i,r}$ that we obtain when analyzing $P_i$ may be
syntactically identical to a term $\tau_{j,s}$ that we obtained when analyzing
some earlier $P_j$.  In such a case, $\tau_{i,r}$ and $\tau_{j,s}$ have
different interpretations.  For example, when analyzing \exref{P1P2}, the two
most important terms are 
$\tau_{1,1} = \prog{g}'-\prog{g} - 1$ and 
$\tau_{2,1} = \prog{g}'-\prog{g} - 1$.  However, $\tau_{1,1}$ represents
the increase to $\prog{g}$ as a result of a call to \textit{P1} and $\tau_{2,1}$
represents the increase to $\prog{g}$ as a result of a call to \textit{P2}.
Our technique will attempt to find distinct bounding functions for these
two terms.

Second, on \lineref{SummaryRec}, we replace the call to the intraprocedural
summarization function $\textit{Summary}(P,\summarycall)$.  In the general case, each
procedure $P_i$ might call every other member of its strongly connected component.
To reduce this analysis step to an intraprocedural-analysis problem, we must
replace every such call with a summary formula.
Therefore, for each $P_i$, the call on the analysis subroutine has the form
$\textit{Summary}(P_i,\summarycallfor{P_1},...,\summarycallfor{P_m})$.  
\textit{Summary} analyzes the body of $P_i$ by replacing each
call to some $P_j$ with the formula $\summarycallfor{P_j}$.  The summary formula
thus produced for $P_i$ is denoted by $\summaryrecfor{P_i}$.

\Lineseqref{SummaryExt}{LastLine} of \algref{candidates} are then
executed for each $P_i$.  On \lineref{SummaryExt}, the formula
$\summaryextractfor{P_i}$ is produced by conjoining $\summaryrecfor{P_i}$ with one
equality constraint for each of the terms $\tau_{i,1},...,\tau_{i,n_i}$, but
not the terms $\tau_{j,q}$ for $j \neq i$.  
On \lineref{WedgeExtract}, the call to $\textit{Abstract}$ has the
form 
$\textit{Abstract}(\summaryextractfor{P_i},
  \bsym{1,1}{h},...,\bsym{m,n_m}{h},\bsym{i,q}{h+1})$.  That is, we look for
inequations that provide a bound on $\bsym{i,q}{h+1}$, which relates to $P_i$
specifically, in terms of all of the height-$h$ bounding functions for 
$P_1,...,P_m$.  For example, in \exref{P1P2}, we find the constraints
$\bsym{1,1}{h+1} \leq 18 \bsym{2,1}{h} + 17$ and 
$\bsym{2,1}{h+1} \leq 2 \bsym{1,1}{h} + 1$.

The next steps of height-based analysis are to find a collection of inequations
that form a stratified recurrence, and to solve that
stratified recurrence (as in \sectref{Height}).
These steps are the same in the case of mutual 
recursion as in the case of a single recursive procedure.  After solving the
recurrence, we obtain a closed-form solution for the subset of the bounding
functions $\bsym{1,1}{h},...,\bsym{m,n_m}{h}$ that appeared in the recurrence.
Let $B_i$ be the set of indices $q$ such that we found a recurrence for
$\bsym{i,q}{h}$.
Then, the procedure summary that we obtain for $P_i$ has the following form:

\begin{align}
  \label{Eq:MutualSummary}
  \exists H .
  \summarydepth{P_i}(H,\sigma)
  \land
  \bigwedge_{q \in B_i}
  [
  \tau_q \leq \bsym{i,q}{H}
  ]
\end{align}

In \exref{P1P2}, the recurrence that we obtain is:
\[
  \begin{bmatrix} \bsym{1,1}{h+1} \\ \bsym{2,1}{h+1} \end{bmatrix} 
  \leq
  \begin{bmatrix}  0 & 18 \\
                   2 &  0 
  \end{bmatrix}
  \begin{bmatrix} \bsym{1,1}{h} \\ \bsym{2,1}{h} \end{bmatrix} 
  +
  \begin{bmatrix} 17 \\ 1 \end{bmatrix}
  \label{Eq:P1P2Rec}
\]
Notice that this recurrence involves an interdependency between the
bounding functions for the increase to $\prog{g}$ in \textit{P1} and \textit{P2}.
Simplified versions of the g bounds found by $\Chora$ for \textit{P1}
and \textit{P2} are $3\cdot 6^{\prog{n}-1}$ and $6^{\prog{n}-1}$,
respectively.

\OnlyTech{The extension of two-region analysis (\sectref{DualHeight}) to the case of
mutual recursion is analogous to the extension of height-based recurrence
analysis.  It can be achieved by combining the changes to height-based
recurrence analysis described in \sectref{DualHeight} with the changes to
height-based recurrence analysis described in this sub-section.}

For each procedure within a strongly connected component $S$ of the
call graph, the algorithm of \sectref{Mutual} needs to be able to identify a
base case (i.e., a set of paths containing no calls to the procedures of $S$).
Some programs contain procedures without such base cases. 
\OnlyPaper{
(For a discussion of an extension to our algorithm that can handle
such programs, see the technical report version of this document
\cite[\S 4.5]{arxiv:BCKR2020}.)
}
\OnlyTech{
\subsection{Equation Systems With Missing Base Cases}
\label{Se:MissingBaseCases}
For each procedure within a strongly connected component $S$ of the
call graph, the algorithm of \sectref{Mutual} needs to be able to identify a
base case (i.e., a set of paths containing no calls to the procedures of $S$).
Some programs contain procedures without such base cases, as in the following example.
\begin{example}
\label{Exa:P3P4}
\[
  \begin{array}{l}
    \hspace{0.0cm} \textrm{void}~P3(\textrm{int}~n)~\{                              \\
    \hspace{0.5cm}     \textbf{if}~(n <= 1)~\{~ P4(n-1);P4(n-1); \textbf{return};\} \\
    \hspace{0.5cm}     P3(n-1);~P4(n-1);                                            \\
    \hspace{0.0cm} \}                                                               \\
  \end{array} \\ 
\]
\[
  \begin{array}{l}
    \hspace{0.0cm} \textrm{void}~P4(\textrm{int}~n)~\{                              \\
    \hspace{0.5cm}     \textbf{if}~(n <= 1)~\{~ cost\textrm{++};~\textbf{return};\} \\
    \hspace{0.5cm}     P4(n-1);~P3(n-1);                                            \\
    \hspace{0.0cm} \}                                                               \\
  \end{array}
\]

Notably, every path through $P_3$ makes a call on either $P_3$ or $P_4$.
When \algref{candidates} is applied to $P_3$, the base case summary
$\summarybasefor{P_3}$ will be the transition formula \textit{false},
because $\summarybasefor{P_3}$ is computed in a way that excludes all paths
containing calls that are potentially indirectly recursive.  
Thus, no bounded terms will be found when analyzing $\summarybasefor{P_3}$.
The procedure-summary equation system for these two procedures is shown below
as \eqref{HardSystem}.  In \eqref{HardSystem}, the variables $P_3$ and $P_4$
stand for the procedure summaries, and $a$ is the base case of
$P_4$, i.e. the action that adds one to the global variable $\prog{cost}$.
\begin{align}
    P_3 &= (P_4 \otimes P_4) \oplus (P_3 \otimes P_4) \hspace{15.0ex} \\
    P_4 &= a \oplus (P_4 \otimes P_3) \label{Eq:HardSystem}
\end{align}
\end{example}

We can solve this problem by transforming the equation system in the 
following manner.  
For each 
$j \in \{1,...,i-1,i+1,...,m\}$, create a new procedure-summary variable 
$P_{j \setminus \{i\}}$ to represent executions of $P_j$ that never result
in a call back to $P_i$.  
Next, replace every call to $P_j$ in the equation for $P_i$ with a call to
$(P_j \oplus P_{j \setminus \{i\}})$ (so that a call to $P_j$ is allowed to
either call back to $P_i$ or not do so).  
Let the original equation for $P_j$ be $P_j = RHS$.
Then, create an equation for $P_{j \setminus \{i\}}$ by
replacing $P_i$ with the trivial summary $0$ (i.e., \textit{abort}) in $RHS$.
Applying this transformation to \eqref{HardSystem} yields:
\begin{align*}
  P_3 &= 
    (P_3 \otimes (P_4 \oplus P_{4 \setminus\{3\}})) \oplus 
    ((P_4 \oplus P_{4 \setminus\{3\}}) \otimes (P_4 \oplus P_{4 \setminus\{3\}})) 
    \\
  P_4 &= (P_4 \otimes P_3) \oplus a \\
  P_{4 \setminus\{3\}} &= 
    a \oplus (P_{4 \setminus\{3\}} \otimes 0) = a
\end{align*}
Observe that $P_{4 \setminus\{3\}}$, considered as a procedure, lies outside
of the call-graph strongly-connected-component $\{P_3,P_4\}$, because it calls
neither $P_3$ nor $P_4$.  Therefore, $P_{4 \setminus\{3\}}$  can be analyzed
using the algorithms of this paper to produce a summary, and we can use that
summary when we return to the analysis of $\{P_3,P_4\}$.  Subsequently, when
we analyze $\{P_3,P_4\}$, we find a base case for $P_3$ corresponding to the
path $P_{4 \setminus\{3\}} \otimes P_{4 \setminus\{3\}}$, which 
corresponds to the action of adding two to $\prog{cost}$.

Each time we apply the above transformation, we create $m-1$
new procedures $P_{j \setminus\{i\}}$ for $j \in \{1,...,i-1,i+1,...,m\}$.
For some equation systems, we must apply this transformation for several
such $i$.
In the worst case, the transformation can lead to a worst-case increase of
$O(2^m)$ in the number of variables in the equation system.
} 



\section{Experiments}
\label{Se:Experiments}

Our techniques are implemented as an interprocedural extension of Compositional
Recurrence Analysis (CRA) \cite{FMCAD:FK15}, resulting in a tool we call
Compositional Higher-Order Recurrence Analysis ($\Chora$).

CRA is a program-analysis tool that uses recurrences to
summarize loops, and uses Kleene iteration to summarize
recursive procedures. 
Interprocedural Compositional Recurrence Analysis (ICRA) \cite{PLDI:KBFR17} is
an earlier extension of CRA that lifts CRA's recurrence-based loop summarization to 
summarize \emph{linearly} recursive procedures.
However, ICRA resorts to Kleene iteration in the case of non-linear recursion.
$\Chora$ can analyze programs containing arbitrary combinations of loops
and branches using CRA.
In the case of linear recursion, $\Chora$ uses the same reduction to CRA as
ICRA.  Thus, in those cases, $\Chora$ will produce results almost identical to
those of ICRA.
The algorithms of \sectref{TechnicalDetails}, which allow $\Chora$ to perform a
precise analysis of non-linear recursion, are what distinguish $\Chora$ from
prior work. 
For this reason, our experiments are focused on the analysis of
non-linearly recursive programs.

Our experimental evaluation is designed to answer the following question:
\begin{eqbox}{Question}
Is $\Chora$ effective at generating invariants for programs containing
non-linear recursion?
\end{eqbox}
Despite the prominence of non-linear recursion (e.g., divide-and-conquer algorithms),
there are few benchmarks in the verification literature that
make use of it.
The examples that we found are bounds-generation benchmarks that come from
the \emph{complexity-analysis} literature, as well as assertion-checking
benchmarks from the \emph{recursive} subcategory of SV-COMP.

\paragraph{Generating complexity bounds.}
For our first set of experiments, we evaluate $\Chora$ on
twelve benchmark programs from the complexity-analysis literature.
This set of experiments is designed to determine how the complexity-analysis
results obtained by $\Chora$ compare with those obtained by ICRA and
state-of-the-art complexity-analysis tools.
We selected all of the non-linearly recursive programs in the benchmark suites
from a recent set of complexity-analysis papers
\cite{TOPLAS:CFG19,PLDI:CHS15,TR:KH19}, as well as the web site of PUBS
  \cite{PUBSwebsite}, and removed duplicate (or near-duplicate)
  programs, and translated them to C.
Our implementations
of divide-and-conquer algorithms are working implementations rather than 
cost models, and therefore $\Chora$'s analysis of these programs involves
performing non-trivial invariant generation and cost analysis at the same time.
Source code for $\Chora$ and all benchmarks can be found in the $\Chora$
repository \cite{CHORAwebsite}.

To perform a complexity analysis of a program using $\Chora$,
we first manually modify the program to add an
explicit variable ($\prog{cost}$) that tracks the time (or some other resource)
used by the program.  
We then use $\Chora$ to generate a term that bounds the final value
of $\prog{cost}$ as a function of the program's inputs.  
Note that, as a consequence of this technique, $\Chora$'s bounds on a program's
running time are only sound under the assumption that the program terminates.
Throughout the analysis, $\Chora$ merely treats $\prog{cost}$ as
another program variable; that is, the recurrence-based analytical
techniques that it uses to perform cost analysis are the same as those
it uses to find all other numerical invariants.

The benchmark programs on which we evaluated $\Chora$, as well as the
complexity bounds obtained by $\Chora$'s analysis, are shown in
\tableref{BoundsTable}.  
The first five programs are elementary examples of non-linear
recursion.  
The next seven are more challenging complexity-analysis problems that have been
used to test the limits of state-of-the-art complexity analyzers.  

\begin{table}
    \centering
    {\small \caption{\label{Ta:BoundsTable}
        {\small 
        Column 2 shows the actual asymptotic bound for each benchmark program. 
        Columns 3-4 show the asymptotic complexity of the bounds determined by
        $\Chora$ and ICRA.  Column 5 gives the source of the benchmark 
        as well as the published bound from that source.
        ``n.b.'' indicates that no bound was found.
        For each benchmark, only one other tool's bound is shown, even if more
        than one such tool is capable of finding a bound.
    }}}
    \resizebox{.48\textwidth}{!}{
        \begin{tabular}{@{\hspace{0ex}}||@{\hspace{.35ex}}l@{\hspace{.35ex}}||@{\hspace{.35ex}}l@{\hspace{.35ex}}||@{\hspace{.35ex}}l@{\hspace{.35ex}}||@{\hspace{.35ex}}l@{\hspace{.35ex}}||@{\hspace{.35ex}}l@{\hspace{.35ex}}||@{\hspace{0ex}}}
\hhline{|t:=====:t|}
                   Benchmark     & Actual               & $\Chora$             & ICRA   & Other Tools \\
\hhline{||=====||} fibonacci     & $O(\varphi^n)$       & $O(2^n)$             & n.b.   & \cite{PUBSwebsite}:$O(2^n)$ \\
\hhline{||-----||} hanoi         & $O(2^n)$             & $O(2^n)$             & n.b.   & \cite{PUBSwebsite}:$O(2^n)$ \\
\hhline{||-----||} subset\_sum   & $O(2^n)$             & $O(2^n)$             & n.b.   & \cite{TR:KH19}:$O(2^n)$ \\
\hhline{||-----||} bst\_copy     & $O(2^n)$             & $O(2^n)$             & n.b.   & \cite{PUBSwebsite}:$O(2^n)$ \\
\hhline{||-----||} ball\_bins3   & $O(3^n)$             & $O(3^n)$             & n.b.   & \cite{TR:KH19}:$O(3^n)$ \\
\hhline{||=====||} karatsuba     & $O(n^{\log_2{(3)}})$ & $O(n^{\log_2{(3)}})$ & n.b.   & \cite{TOPLAS:CFG19}:$O(n^{1.6})$ \\
\hhline{||-----||} mergesort     & $O(n \log(n))$       & $O(n \log(n))$       & n.b.   & \cite{PUBSwebsite}:$O(n \log(n))$ \\
\hhline{||-----||} strassen      & $O(n^{\log_2{(7)}})$ & $O(n^{\log_2{(7)}})$ & n.b.   & \cite{TOPLAS:CFG19}:$O(n^{2.9})$ \\
\hhline{||-----||} qsort\_calls  & $O(n)$               & $O(2^n)$             & $O(n)$ & \cite{PLDI:CHS15}:$O(n)$ \\
\hhline{||-----||} qsort\_steps  & $O(n^2)$             & $O(n 2^n)$           & n.b.   & \cite{TOPLAS:CFG19}:$O(n^{2})$ \\ 
\hhline{||-----||} closest\_pair & $O(n \log(n))$       & n.b.                 & n.b.   & \cite{TOPLAS:CFG19}:$O(n \log(n))$ \\
\hhline{||-----||} ackermann     & $Ack(n)$             & n.b.                 & n.b.   & \cite{PUBSwebsite}:n.b. \\
\hhline{|b:=====:b|}
        \end{tabular}
    } 
\end{table}

We observe that on two benchmarks, \benchmark{karatsuba} and
\benchmark{strassen}, $\Chora$ finds an asymptotically tight bound
that was not found by the technique from which the benchmark was
taken.
For example, the bound obtained by $\Chora$ for \benchmark{karatsuba} has the form
$\prog{cost} \leq 3^{\log_2(n)}$ which is equivalent to $\prog{cost} \leq
n^{\log_2(3)}$, and is therefore tighter than the bound using the rational
exponent 1.6 cited in \cite{TOPLAS:CFG19}, although the technique from \cite{TOPLAS:CFG19} can
obtain rational bounds that are arbitrarily close to $\log_2(3)$.
On two benchmarks, $\Chora$ fails to produce an asymptotically tight bound.
For example, for \benchmark{qsort\_steps}, \prog{cost} tracks the number of instructions,
$\Chora$ finds an exponential bound (as does the PUBS complexity analyzer
\cite{PUBSwebsite}, which also uses recurrence solving and height-based
abstraction), whereas \cite{TOPLAS:CFG19} finds the optimal $O(n^2)$ bound.
On two more benchmarks, $\Chora$ is unable to find a bound.
Note that $\Chora$'s technique for summarizing recursive functions
significantly improves upon ICRA's, which can find only one bound across the suite.

\begin{figure}[t]
\vspace{-2.0ex}
\includegraphics[width=0.48\textwidth]{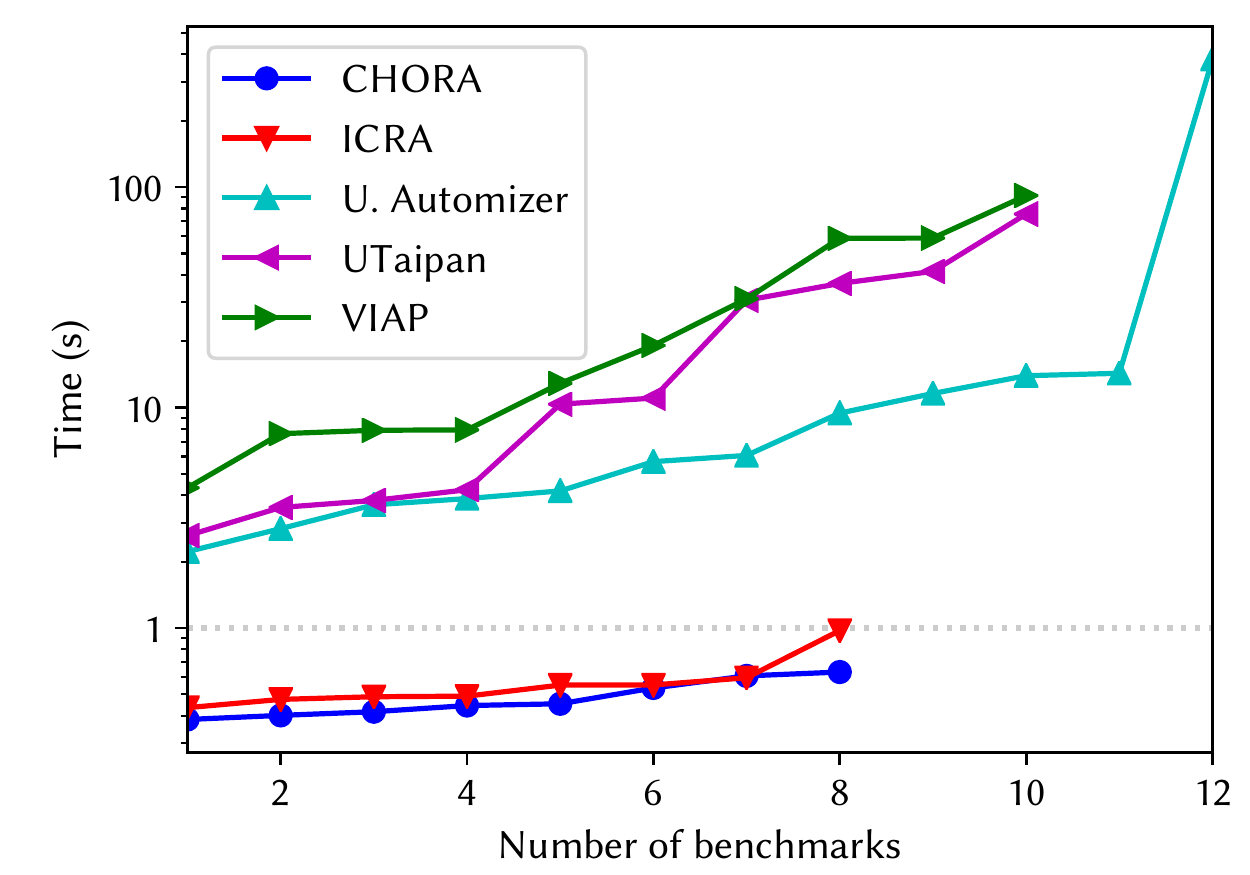} 
\caption{\label{Fi:Cactus}
{\small Results of running CHORA and four other tools on the SV-COMP19
\textit{recursive} directory of benchmarks.  Each point indicates a benchmark
containing assertions that a tool proved to be true, and the amount of
time taken by that tool on that benchmark.}
}
\end{figure}

\paragraph{Assertion-checking experiments.}
Next, we tested $\Chora$'s invariant-generation abilities on
assertion-checking benchmarks.  A standard benchmark suite from the literature
is the Software Verification Competition (SV-COMP), which includes a recursive
sub-category (\emph{ReachSafety-Recursive}).
Within this sub-category, we selected the benchmarks in the \emph{recursive}
sub-directory that contained true assertions, yielding a set of 17 benchmarks.
We ran $\Chora$, ICRA, and the top three performers on this category from the
2019 competition: Ultimate Automizer (UA) \cite{TACAS:HCDEHLNSP13},
UTaipan \cite{TACAS:DGHHNPSS2018}, and VIAP \cite{SYNASC:RL2017}.  
\figref{Cactus} presents a cactus plot showing the number of benchmarks proved
by each tool, as well as the timing characteristics of their runs.

Timings were taken on a virtual machine running Ubuntu 18.04 with 16 GB
of RAM, on a host machine with 32GB of RAM and a 3.7 GHz Intel i7-8000K CPU.  
These results demonstrate that $\Chora$ is roughly
an order of magnitude faster for each benchmark than the other tools.
UA proved the assertions in 12 out of 17 benchmarks; UTaipan and VIAP each
proved the assertions in 10 benchmarks;
$\Chora$ proved the assertions in 8 benchmarks; 
all other tools from the competition proved the assertions in 6 or fewer
benchmarks.

\begin{figure}[t]
\[
{ \small
\begin{array}{l}
\hspace{0.0cm} \textrm{int}~\textit{ackermann}(\textrm{int}~m,\textrm{int}~n)~\{                    \\ 
\hspace{0.5cm}     \textbf{if}~(m == 0)~\{~\textbf{return}~n+1;\}                                   \\
\hspace{0.5cm}     \textbf{if}~(n == 0)~\{~\textbf{return}~\textit{ackermann}(m-1,1);\}             \\
\hspace{0.5cm}     \textbf{return}~\textit{ackermann}(m-1,\textit{ackermann}(m,n-1));               \\
\hspace{0.0cm} \}                                                                                   \\
\hspace{0.0cm} \textit{assert}(n < 0~||~m < 0~||~\textit{ackermann}(m,n) >= 0)                      \\
\hline                                                                                           
\hspace{0.0cm} \textrm{int}~\textit{hanoi}(\textrm{int}~n)~\{                                    \\ 
\hspace{0.5cm}     \textbf{if}~(n == 1)~\{~\textbf{return}~1;\}                                  \\
\hspace{0.5cm}     \textbf{return}~2*(\textit{hanoi}(n-1)) + 1;                                  \\
\hspace{0.0cm} \}                                                                                \\
\hspace{0.0cm} \textrm{void}~\textit{applyHanoi}(\textrm{int}~n,\textrm{int}~from,\textrm{int}~to,\textrm{int}~via)~\{ \\
\hspace{0.5cm}     \textbf{if}~(n == 0)~\{~\textbf{return};\}                                    \\
\hspace{0.5cm}     counter\texttt{++};                                                           \\
\hspace{0.5cm}     \textit{applyHanoi}(n-1,from,via,to);                                         \\
\hspace{0.5cm}     \textit{applyHanoi}(n-1,via,to,from);                                         \\
\hspace{0.0cm} \}                                                                                \\
\hspace{0.0cm} counter = 0; \textit{applyHanoi}(n,...);~\textit{assert}(\textit{hanoi}(n) == counter) \\
\hline
\hspace{0.0cm} \textrm{int}~\textit{f91}(\textrm{int}~\textit{x})~\{                            \\ 
\hspace{0.5cm}     \textbf{if}~(x > 100)~\textit{return}~x-10; 
                   \textbf{else}~\{~\textit{return}~\textit{f91}(\textit{f91}(x+11))~\};        \\
\hspace{0.0cm} \}                                                                               \\
\hspace{0.0cm} res = f91(x); \textit{assert}(res==91~||~x > 101 ~\&\&~ res == x - 10) \\
\end{array}
}
\]
\caption{\label{Fi:SVCOMPcode} \small
Source code for three programs from the SV-COMP suite: Ackermann01, RecHanoi01, and McCarthy91}
\end{figure}

While the SV-COMP benchmarks do give some insight into $\Chora$'s
invariant-generation capability, the recursive suite is not an ideal test of that
capability, because the suite contains many benchmarks that can be proved safe
by unrolling (e.g., verifying that Ackermann's function evaluated at (2,2) is
equal to 7).  That is, many of these benchmarks do not actually require an
analyzer to perform invariant generation.  

We now discuss three benchmarks from the SV-COMP suite that do give some
insight into $\Chora$'s capabilities, in that they are non-linearly recursive
benchmarks that require an analyzer to perform invariant-generation.
The Ackermann01 benchmark contains an implementation of the two-argument
Ackermann function, and the benchmark asserts that the return value of Ackermann
is non-negative if its arguments are non-negative; $\Chora$ is able to prove
that this assertion holds.  
The RecHanoi01 benchmark contains a non-linearly recursive cost-model of the
Tower of Hanoi problem, along with a linearly recursive function that doubles
its return value and adds one at each recursive call.  
The assertion in
recHanoi01 states that these two functions compute the same value, and $\Chora$
is able to prove this assertion.  
(The other tools that we tested, namely ICRA, UA, UTaipain, and VIAP, were not
able to prove this assertion.)
The McCarthy91 benchmark contains an implementation of McCarthy's 91 function,
along with an assertion that the return value of that function, when applied to
an argument $x$, either (1) equals 91, or else (2) equals $x - 10$.
$\Chora$ is not well-suited to prove this assertion because the asserted
property is a disjunction, i.e., it describes the return value using two
cases, whereas the hypothetical summaries used by $\Chora$ do not contain
disjunctions.
(ICRA, UA, UTaipan, and VIAP were all able to prove this assertion.)

To further test $\Chora$'s capabilities, we also manually created three new
assertion-checking benchmarks, shown in \figref{AssertCode}.  
Because our goal is to assess $\Chora$'s ability to synthesize invariants, our
additional suite consists of recursive examples for which unrolling is an
impractical strategy.

\begin{figure}[t]
\[
{ \small
\begin{array}{l}
\hspace{0.0cm} \textrm{int}~\textit{quad}(\textrm{int}~m)~\{                       \\ 
\hspace{0.5cm}     \textbf{if}~(m == 0)~\{~\textbf{return}~0;\}                    \\
\hspace{0.5cm}     \textrm{int}~\textit{retval};                                   \\
\hspace{0.5cm}     \textbf{do}~\{~\textit{retval}=\textit{quad}(m-1)+m~\}~\textbf{while}(*);        \\
\hspace{0.5cm}     \textbf{return}~\textit{retval};                                \\
\hspace{0.0cm} \}                                                                  \\
\hspace{0.0cm} \textit{assert}(\textit{quad}(n)*2==n+n*n)                          \\
\hline
\hspace{0.0cm} \textrm{int}~\textit{pow2\_overflow}(\textrm{int}~p)~\{                     \\ 
\hspace{0.5cm}     \textit{// \textit{pow2\_overflow} is called with $0 \leq p \leq 29$}   \\
\hspace{0.5cm}     \textbf{if}~(p == 0)~\{~\textbf{return}~1;\}                            \\
\hspace{0.5cm}     \textrm{int}~\textit{r1}=\textit{pow2\_overflow}(p-1);                  \\
\hspace{0.5cm}     \textrm{int}~\textit{r2}=\textit{pow2\_overflow}(p-1);                  \\
\hspace{0.5cm}     \textit{assert}(\textit{r1} + \textit{r2} < 1073741824);                \\
\hspace{0.5cm}     \textbf{return}~\textit{r1}+\textit{r2};                                \\
\hspace{0.0cm} \}                                                                          \\
\hline
\hspace{0.0cm} \textrm{int}~\textit{height}(\textrm{int}~\textit{size})~\{                        \\ 
\hspace{0.5cm}     \textbf{if}~(\textit{size} == 0)~\{~\textbf{return}~0;\}                       \\
\hspace{0.5cm}     \textrm{int}~\textit{left\_size}=\textit{nondet}(0,\textit{size});~//~0 \leq \textit{left\_size} < \textit{size} \\
\hspace{0.5cm}     \textrm{int}~\textit{right\_size}=\textit{size}-\textit{left\_size}-1;         \\
\hspace{0.5cm}     \textrm{int}~\textit{left\_height}=\textit{height}(\textit{left\_size});       \\
\hspace{0.5cm}     \textrm{int}~\textit{right\_height}=\textit{height}(\textit{right\_size});     \\
\hspace{0.5cm}     \textbf{return}~1+\textit{max}(\textit{left\_height},\textit{right\_height});  \\
\hspace{0.0cm} \}                                                                                 \\
\hspace{0.0cm} \textit{assert}(height(n) \leq n)                                                  \\
\end{array}
}
\]
\caption{\label{Fi:AssertCode}
Source code for three non-linearly recursive programs containing assertions.
}
\end{figure}

\benchmark{quad} has a recursive call in a loop that may run for
arbitrarily many iterations, and its return value is always
$\prog{n}(\prog{n}+1)/2$.
\benchmark{pow2\_overflow} contains an assertion inside a non-linearly
recursive function, and an assumption about the range of parameter values;
if the assertion passes, we may conclude that the program is safe from
numerical-overflow bugs.
The benchmark \benchmark{height} asserts that the size (i.e., the number
of nodes) of a tree of recursive calls is an upper bound on the height
of the tree of recursive calls.

\newcommand{\acyes}{\checkmark}
\newcommand{\acno}{\text{\sffamily X}}
\begin{table}
    \centering
    {\small \caption{\label{Ta:AssertTable}
        {\small Five analysis tools, along with the results of
        assertion-checking experiments using the benchmarks shown in
        \figref{AssertCode}.  A $\acyes$ indicates that the tool was able to
        prove the assertion within 900 seconds, and an $\acno$ indicates that
        it was not.  We also show the time required to analyze each benchmark.
    }}}
    \resizebox{.48\textwidth}{!}{
        \begin{tabular}{@{\hspace{0ex}}||@{\hspace{.35ex}}l@{\hspace{.35ex}}||@{\hspace{.35ex}}l@{\hspace{.35ex}}||@{\hspace{.35ex}}l@{\hspace{.35ex}}||@{\hspace{.35ex}}l@{\hspace{.35ex}}||@{\hspace{.35ex}}l@{\hspace{.35ex}}||@{\hspace{.35ex}}l@{\hspace{.35ex}}||@{\hspace{0ex}}}
\hhline{|t:======:t|} 
                    Benchmark                   & $\Chora$       & ICRA           & UA             & UTaipan         & VIAP         \\
\hhline{||======||} \benchmark{quad}            & \acyes (0.70s) & \acyes (1.08s) & \acno  (900s)  & \acyes  (4.24s) & \acno (4.71s)\\     
\hhline{||------||} \benchmark{pow2\_overflow}  & \acyes (0.61s) & \acyes (1.28s) & \acno  (900s)  & \acno   (900s)  & \acno (1.79s)\\ 
\hhline{||------||} \benchmark{height}          & \acyes (0.58s) & \acno  (0.52s) & \acyes (8.82s) & \acyes  (13.0s) & \acno (2.85s)\\
\hhline{|b:======:b|}
        \end{tabular}
    } 
\end{table}

The results of our experiments are shown in \tableref{AssertTable}.
$\Chora$ is able to prove the assertions in all three programs; ICRA and
UTaipan each prove two; UA proves one, and VIAP proves none.
Times taken by each tool are also shown in the table.
$\Chora$'s ability to prove the assertion in \benchmark{quad}
illustrates that it can find invariants even for programs in which 
running time (and the number of recursive calls) is unbounded.
\benchmark{quad} illustrates $\Chora$'s applicability to perform
program-equivalence tasks on numerical programs,
while \benchmark{pow2\_overflow} illustrates
$\Chora$'s applicability to perform overflow-checking.

\paragraph{Conclusions.}
Our main experimental question is whether $\Chora$ is effective at the
problem of generating invariants for programs using non-linear recursion.
Results from the complexity-analysis and assertion-checking experiment show
that $\Chora$ is able to generate non-linear invariants that are sufficient
to solve these kinds of problems.
In these ways, $\Chora$ has shown success in a domain, i.e., invariant
generation for non-linearly recursive programs, that is not
addressed by many other tools.



\section{Related Work}
\label{Se:Related}

Following the seminal work of Cousot and Cousot \cite{POPL:CC77}, most
invariant-generation techniques are based on \textit{iterative
  fixpoint computation}, which over-approximates Kleene-iteration
within some abstract domain.  This paper presents a
\textit{non-iterative} method for generating numerical invariants for
recursive procedures, which is based on extracting and solving recurrence
relations. 
It was inspired by two streams of ideas found in prior work.

\smallskip
\noindent
\textbf{Template-based methods} fix a desired template for the
invariants in a program, in which there are undetermined constant
symbols \cite{CAV:CSS03,SAS:SSM04}.  Constraints on the constants are
derived from the structure of the program, which are given to a
constraint solver to derive values for the constants.  The
\textit{hypothetical summaries} introduced in \sectref{Height} were
inspired by template-based methods, but go beyond them in an important
way: in particular, the indeterminates in a hypothetical summary are
\textit{functions} rather than constants, and our work uses recurrence
solving to synthesize these functions.

Of particular relevance to our work are template-based methods for
generating non-linear invariants
\cite{Kapur:ACA2004,POPL:SSM2004,SAS:CJJK2012,SAS:KKS2016,TOPLAS:CFG19}.
Contrasting with the technique proposed in this paper, a distinct
advantage of template-based methods for generating polynomial
invariants for programs with real-typed variables is that they enjoy
completeness guarantees
\cite{Kapur:ACA2004,POPL:SSM2004,TOPLAS:CFG19}, owing to the
decidability of the theory of the reals.  The advantages of our
proposed technique over traditional template-based techniques are (1)
it is compositional, (2) it can generate exponential and logarithmic
invariants, and (3) it does not require fixing bounds on polynomial
degrees \textit{a priori}.  Also note that template-based techniques
pay an up-front cost for instantiating templates that is exponential
in the degree bound. (In practice, this exponential blow-up can be
mitigated \cite{SAS:KKS2016}.)

\smallskip
\noindent
  \textbf{Recurrence-based methods} find loop invariants
  by extracting recurrence relations between the pre-state and
  post-state of the loop and then generating invariants from their
  closed forms
  \cite{FMCAD:FK15,PACMPL:KCBR18,POPL:KBCR19,ISAAC:RCK2004,ISAAC:HJK17,TACAS:Kovacs2008,ATVA:OBP2016,VMCAI:HJK2018}.
  This paper gives an answer to the question of how such analyses can
  be applied to recursive procedures rather than loops, by extracting
  height-indexed recurrences using template-based techniques.

  \citet{TOPLAS:RTP17} demonstrate that tensor products can be used to
  apply loop analyses to \textit{linearly} recursive procedures.  This
  technique is used in the recurrence-based invariant generator ICRA
  to handle linear recursion \cite{PLDI:KBFR17}.  ICRA falls back on a
  fixpoint procedure for non-linear recursion; in contrast, the
  technique presented in this paper uses recurrence solving to analyze recursive
  procedures.

\citet{SYNASC:RL2017} presents a verification technique that analyzes
recursive procedures by encoding them into first-order logic;
recurrences are extracted and replaced with closed forms as a
simplification step before passing the query to a theorem prover.  In
contrast to this paper, \citet{SYNASC:RL2017}'s approach has the
flexibility to use other approaches (e.g., induction) when
recurrence-based simplification fails, but cannot be used for
general-purpose invariant generation.

\smallskip
\noindent
\textbf{Resource-bound analysis} \cite{CACM:Wegbreit1975} is another related
area of research. 
Three lines of recent research in resource-bound analysis are
represented by the tools PUBS \cite{JAR:AAG2011}, CoFloCo
\cite{Thesis:FloresMontoya}, KoAT \cite{TOPLAS:BEFFG2016}, and
RAML \cite{CAV:HAH12}.
In resource-bound analysis, the goal is to find an expression that
upper-bounds or lower-bounds the amount of some resource (e.g., time,
memory, etc.) used by a program.  Resource-bound analysis typically
consists of two parts: (i) \textit{size analysis}, which finds
invariants that bound program variables, and (ii) \textit{cost
  analysis}, which finds bounds on cost using the results of the size
analysis.  Cost can be seen as an auxiliary program variable, although
it is updated in a restricted manner (by addition only), it has no
effect on control flow, 
and it is often assumed to be non-negative.
Our work differs from resource-bound analyzers in several ways, ultimately
because our goal is to find invariants and check assertions, rather than to
find resource bounds specifically.

The capabilities of our technique are different, in that we are able to find
non-linear mathematical relationships (including polynomials, exponentials,
and logarithms) between variables, even in non-linearly recursive procedures.
PUBS and CoFloCo use polyhedra to represent invariants, so they are
restricted to finding linear relationships between variables, although they can
prove that programs have non-linear costs.  
KoAT has the ability to find non-linear (polynomial and exponential) bounds on
the values of variables, but it has limited support for analyzing non-linearly
recursive functions; in particular, KoAT cannot reason about the transformation
of program state performed by a call to a non-linearly recursive function.  
Typically, resource-bound analyzers also reason about non-terminating
executions of a program, whereas our analysis does not.
RAML reasons about manipulations of data structures, whereas our work
only reasons about integer variables.  Originally, RAML only discovered
polynomial bounds, although recent work \cite{TR:KH19} extends the technique to
find exponential bounds.

The algorithms that we use are different in that we have a unified approach, 
rather than separate approaches, for analyzing cost and analyzing a program's
transformation of other variables.  To perform resource-bound analysis, we
materialize cost as a program variable and then find a procedure summary; the
summary describes the program's transformation of all variables,
including the cost variable.
Recurrence-solving is
the essential tool that we use for analyzing 
loops, linear recursion, and non-linear recursion,
and 
we are able to find non-linear mathematical relationships because such
relationships arise in the solutions of recurrences.


\begin{acks}                            
Supported, in part,
by a gift from Rajiv and Ritu Batra;
by \grantsponsor{GS100000003}{ONR}{https://www.onr.navy.mil/} under grants
  ~\grantnum{GS100000003}{N00014-17-1-2889} and
  ~\grantnum{GS100000003}{N00014-19-1-2318}.
The U.S.\ Government is authorized to reproduce and distribute
reprints for Governmental purposes notwithstanding any copyright
notation thereon.
Opinions, findings, conclusions, or recommendations
expressed in this publication are those of the authors,
and do not necessarily reflect the views of the sponsoring
agencies.
\end{acks}

\bibliography{references}

\OnlyTech{
\appendix

\newpage
\section{Appendix}
\label{Se:Appendix}

\newcommand{\assumep}[1]{\textit{assume}(#1)}
\newcommand{\aug}{\mathcal{M}}
\newcommand{\Paug}{\hat{P}}
\newcommand{\taug}{\hat{t}}

In this section, we provide a detailed argument for the soundness of 
height-based recurrence analysis.
We discuss the process of performing a height-based recurrence analysis
on some procedure $P$.  The sequence of operations in that analysis is as
follows.
First, \algref{candidates} analyzes $P$, and produces as output
a set of candidate recurrence inequations. 
On \lineseqref{GetBase}{CreateSymbols}, \algref{candidates} also produces a set
$\{\tau_i\}_{i \in [1,n]}$ of two-vocabulary relational expressions.
Next, \algref{filtering} filters down the set of candidate recurrences produced
by \algref{candidates} to obtain a stratified recurrence that can be solved by
a C-finite recurrence solver.  Finally, a recurrence solver produces
a solution in the form of a set of functions $\{b_i\}_{i \in B}$, where
$B=\{i_1,...,i_m\}$ is a subset of the indices $[1,n]$.

In this discussion of soundness, we wish to relate the functions $\{b_i\}_{i
\in B}$ that are produced by the analysis to the sets of values $V_\tau(P,h)$
taken on by each relational expression $\tau_i$ at each height $h$, which have
the following definition in terms of the relational semantics given in
\sectref{Background}:
\[ V_\tau(P,h) \eqdef \{ \expsem{\tau}(\sigma,\sigma') : \tuple{\sigma,\sigma'} \in R(P,h) \}. \]
We use $V_\tau(P,h)$ to prove a height-relative soundness property of our
procedure summaries, which contain the height $h$ as an explicit parameter.
The fact that the summaries contain an explicit representation of height
means that they can be made more precise by conjoining them to the depth-bound
summaries computed in \sectref{DepthBound}.

The goal of this section is to prove the following soundness theorem.
\begin{theorem}
\label{The:Soundness}
Let $P$ be a procedure to which \algref{candidates} and \algref{filtering} have
been applied to obtain stratified recurrence.
Let $\{\tau_i\}_{i \in [1,n]}$ be the relational expressions computed by
\algref{candidates}.
Let $B \subseteq [1,n]$ be such that $\{b_i\}_{i \in B}$ is 
the set of functions produced by solving the stratified recurrence.
Then, the following statement holds:
$\forall h \geq 1 . \bigwedge_{i \in B} \forall v \in V_{\tau_i}(P,h). v \leq b_i(h)$.
\end{theorem}
We will prove \theoref{Soundness} by induction on the height $h$.  
However, before the main inductive argument, we will provide some definitions,
and discuss the properties of \algref{candidates}, the recurrence-extraction
algorithm \algref{filtering}, and the set of functions $\{b_i\}_{i \in B}$.
(Note that the \algref{filtering} referred to in this appendix is not the same
as the Alg.~3 that appears in the conference version \cite{PLDI:BCKR2020} of
this document, which appears as the depth-bounding algorithm \algref{depth} in
this technical report version.)

Define a \textit{feasible trace} of a procedure $P$ to be a finite
list of pairs of control locations and program states, starting at the entry
location of $P$, ending at the exit location of $P$, in which all the state
transitions are consistent with the semantics of $P$, and all calls are matched
by returns.  Note that this definition only considers finite (i.e.,
terminating) traces of $P$, which is useful because our ultimate
goal is to find procedure summaries that over-approximate a procedure's
pre-state-post-state relation, and a procedure only has a post-state when it
terminates.  
(As described below, we will also discuss a modified version of $P$ called
$\Paug$, and we will consider the \textit{feasible traces of} $\Paug$ to be 
only those that meet some additional constraints.)
Furthermore, we define the \textit{feasible traces of} $P$ \textit{up to height}
$h$ to be those feasible traces that have a recursion depth not exceeding $h$.
For the following soundness proof, we define \textit{invariants of}
$P$ to be properties that hold in all feasible traces of $P$.  

As explained above, the main function of \algref{filtering} is to filter down
the set of candidate recurrence inequations produced by \algref{candidates},
to obtain a subset that constitute a stratified recurrence.  At the end of
that process, the inequations are changed into equations so as to obtain the
maximal solution to the set of inequations.
The output of \algref{filtering} is a stratified recurrence that can be written as:
\[
  \bigwedge_{i \in B} b_{i}(h+1) = p_i(b_{i_1}(h),\ldots,b_{i_m}(h)),
\]
in which each $p_i(x_1,\ldots,x_m)$ is a polynomial in the variables 
$x_1,\ldots,x_m$.  

All coefficients in the polynomials $\{p_i\}_{i \in B}$ are non-negative,
including the constant coefficients, as a result of \lineref{DropNegative} of
\algref{filtering}, which drops terms having negative coefficients from the
polynomial inequations that are given as input to \algref{filtering}, thereby
weakening the inequations.
Aside from \lineref{DropNegative},
all other steps of \algref{filtering} serve only to filter down the set of
candidate recurrences.  Because of the dropping of terms having negative
coefficients on \lineref{DropNegative}, the polynomials $p_i$ may differ from
the corresponding polynomials that appeared in the input to \algref{filtering};
for each $i \in B$, we denote by $p_i'$ the corresponding polynomial in the
input, before terms having negative coefficients were dropped.  
We refer to the candidate inequations involving the $p_i'$ polynomials as the
\textit{selected candidate inequations}, because they are the ones that are
selected by \algref{filtering} for inclusion in the stratified recurrence
(after their terms having negative coefficients are dropped).

During the recurrence-solving phase, the zero vector is used as the initial
condition of the recurrence.  Thus, the set of functions that occur as the
solution to the recurrence satisfy $\forall i \in B, b_i(1)=0$.  
Because all
polynomials $p_i$ have only non-negative coefficients, the functions 
$\{b_i(h)\}_{i \in B}$ are non-negative and non-decreasing for $h \geq 1$.

Our final digression before proving \theoref{Soundness} is a discussion
of \algref{candidates}.
\algref{candidates} operates by manipulating formulas that include a set of
function symbols named $\{b_i(h)\}_{i \in [1,n]}$ and $\{b_i(h+1)\}_{i \in [1,n]}$.
The names of these symbols are the same as those of the corresponding
functions that are derived by recurrence solving; however, in this proof, we
will use separate names for the symbols manipulated by \algref{candidates} and
the corresponding functions.  
Instead of $b_i(h)$ we will refer to the symbol $x_i$,
and instead of $b_i(h+1)$, we will refer to the symbol $y_i$.
Furthermore, although \algref{candidates}, as written, manipulates formulas
that are augmented with these additional symbols, it will be convenient in the
following argument to take an alternative, but equivalent, view, according to which
\algref{candidates} analyzes a modified version of the procedure $P$ called $\Paug$
that is obtained by making three changes to $P$.  

First, $\Paug$ is augmented with a set of immutable auxiliary variables named
$x_1,...,x_n,y_1,...,y_n$.  
Second, we impose a constraint on the feasible traces of $\Paug$,
namely that the pre-state $\sigma$ and post-state $\sigma^\prime$ of any
trace $t$ of $\Paug$ must satisfy the constraint 
$\bigwedge_{i=1}^{n} (y_i = \expsem{\tau_i}(\sigma,\sigma'))$,
or else $t$ is not considered to be a \textit{feasible} trace of $\Paug$.
This constraint is the equivalent of the formula-manipulation performed by
\lineref{SummaryExt} of \algref{candidates}.

Third, the recursive call sites in $P$ are replaced with control-flow edges
that havoc their post-state $\sigma^\prime$ and execute
$\assumep{\summarycall}$.  
That is, the feasible executions of these control-flow edges of $\Paug$ are all
those in which the pre-state $\sigma$ and post-state $\sigma^\prime$ of the
control-flow edge satisfy
$\bigwedge_{i=1}^{n} (\expsem{\tau_i}(\sigma,\sigma') \leq x_i \land x_i \geq 0)$.
The effect of replacing call edges of $P$ in this way is equivalent to
that of the formula-manipulation performed by \lineref{SummaryCall} of
\algref{candidates}.

Having described the above constraints on feasible executions of $\Paug$, we
can now succinctly describe the output of \algref{candidates}: each of the
candidate recurrence inequations returned by \algref{candidates} is an
invariant of $\Paug$, i.e., a property that holds in all feasible executions
of $\Paug$.  
The soundness of these invariants follows from the soundness of the underlying
program-analysis primitives used by \algref{candidates}.
For the proof of \theoref{Soundness}, the crucial invariant of $\Paug$ is the
conjunction of the selected candidate inequations: 
\begin{equation}
  \{ y_i \leq p_i^{\prime}(x_{i_1},...,x_{i_m}) \}_{i \in B}. \label{Eq:PPrimes}
\end{equation}

We now begin the inductive proof of \theoref{Soundness}.

\begin{proof}
The base case of the proof corresponds to a height value of 1, which
in turn corresponds to executions of the base case of procedure $P$.
At height 1, we must show:
\[ \bigwedge_{i \in B} \forall v \in V_{\tau_i}(P,1). v \leq b_i(1), \]
which holds because each relational expression $\tau_i$ was constructed to be
bounded above by zero in the base case, and each bounding function $b_i$
evaluates to zero at height 1.

The inductive step of the proof is as follows.
The inductive hypothesis states that, for some $h$,
\[ \bigwedge_{i \in B} \forall v \in V_{\tau_i}(P,h). v \leq b_i(h), \]
and the goal is to prove that
\[ \bigwedge_{i \in B} \forall v \in V_{\tau_i}(P,h+1). v \leq b_i(h+1). \]
Let $t$ be any feasible trace of $P$ at height up to $h+1$.
We will show that we can modify the trace $t$ to produce a new trace $\taug$,
and we then prove that $\taug$ is a feasible trace of $\Paug$.

To construct $\taug$, first modify $t$ to add the immutable auxiliary
variables $x_1,...,x_m,y_1,...,y_m$ to each program state in $t$.  We choose
the values of these auxiliary variables as follows.  Let $\sigma$ and $\sigma'$
be, respectively, the initial and final states of $t$.  
For each $i \in [1,n]$, set $y_i$ to be the result of evaluating the
relational expression $\tau_i$ using the state pair $(\sigma,\sigma')$, that
is, $y_i = \expsem{\tau_i}(\sigma,\sigma')$.
Define the outermost recursive calls in the feasible trace $t$ to be the
recursive calls to $P$ that do not occur inside any other recursive call to
$P$.  Any feasible trace $t$ is of finite length, and therefore $t$ contains
some finite number of outermost recursive calls.  Thus, the set $R_{i,t}$ of
values taken on by $\tau_i$ evaluated at the pre-state/post-state pairs of each
outermost recursive call in $t$ is a finite set, and therefore we may define
$M_{i,t} \defeq \max(0,\max(R_{i,t}))$ to be the maximum value of $\tau_i$
occurring at any outermost recursive call in $t$.  Now set each of the
auxiliary variables $x_i$ as follows:
\[
    x_i = \begin{cases}
            b_i(h)  & \text{if}~i \in B \\
            M_{i,t} & \text{otherwise}
          \end{cases}
\]
Finally, modify $\taug$ by collapsing all of the intermediate steps of each
outermost recursive call in $t$ into a single state transition, so as to match
the replacement of recursive-call edges of $P$ with their corresponding edges
in $\Paug$.

We now argue that $\taug$ meets all the necessary constraints to be considered
a feasible trace of $\Paug$.  
The constraint on the initial and final state of $\taug$ is that
$\bigwedge_{i=1}^{n} (y_i = \tau_i)$, which holds by construction.
The constraint at each outermost recursive call of $t$ is that, if
$(\sigma,\sigma')$ are, respectively, the pre-state and the post-state of the
call, then
$\bigwedge_{i=1}^{n} (\expsem{\tau_i}(\sigma,\sigma') \leq x_i \land x_i \geq 0)$.
For $i \notin B$, $x_i=M_{i,t}$, and each such $M_{i,t}$ satisfies the constraint
by construction.  

For $i \in B$, $x_i = b_i(h)$.  We must show that $x_i$ is greater than or equal
to $\expsem{\tau_i}(\sigma,\sigma')$ and also greater than or equal to zero.
Each $x_i$ is non-negative because each $b_i(h)$ is non-negative.
By hypothesis, $t$ is a feasible execution trace of $P$ at height $h+1$.  Thus,
each outermost recursive call in $t$ corresponds to an execution of $P$ of height
at most $h$.  
(Note that, if a trace is at height exactly $h+1$, one of its recursive calls
must be at height exactly $h$, but the others may be at any height between 1
and $h$ (inclusive).)
Thus, the inductive hypothesis, \[ \bigwedge_{i \in B} \forall v \in
V_{\tau_i}(P,h). v \leq b_i(h), \] implies that the constraint relating $x_i$
to the value of $\tau_i$ is met at each call.  We conclude that the relevant
constraints on $\taug$ are met, and therefore $\taug$ is a feasible trace of
$\Paug$.

As noted above, the output \algref{candidates} is a set of invariants of $\Paug$,
i.e., properties that hold in all feasible traces of $\Paug$.  
One such property is the conjunction of the selected candidate inequations
shown in \eqref{PPrimes}.
Let $i \in B$.
We conclude that the $i^\textrm{th}$ selected candidate inequation holds in $\taug$:
\[ y_i \leq p_i^{\prime}(x_{i_1},...,x_{i_m}) \]
Let $\sigma$ and $\sigma'$ be the initial and final states of $\taug$.
By the construction of $\taug$, we know that $\expsem{\tau_i}(\sigma,\sigma') = y_i$.
Thus,
\[ \expsem{\tau_i}(\sigma,\sigma') \leq p_i^{\prime}(x_{i_1},...,x_{i_m}) \]
By the construction of $\taug$, we know that, for each $k \in B$, $x_k = b_k(h)$.  Thus,
\[ \expsem{\tau_i}(\sigma,\sigma') \leq p_i^{\prime}(b_{i_1}(h),...,b_{i_m}(h)) \]
As noted above, each $b_i(h)$ is non-negative for any $h \geq 1$.
Thus, because $p_i^\prime$ and $p_i$ are being evaluated at non-negative arguments,
and $p_i$ was derived from $p_i'$ by dropping negative coefficients, we conclude that
$p_i^{\prime}(b_{i_1}(h),...,b_{i_m}(h))
\leq
p_i(b_{i_1}(h),...,b_{i_m}(h))$ and therefore,
\begin{equation}
 \expsem{\tau_i}(\sigma,\sigma') \leq p_i(b_{i_1}(h),...,b_{i_m}(h)).
 \label{Eq:CanUseRec}
\end{equation}
The right-hand side \eqref{CanUseRec} matches the right-hand side of the defining 
recurrence for $b_i$, i.e., $p_i(b_{i_1}(h),...,b_{i_m}(h)) = b_i(h+1)$.  Thus,
\[ \expsem{\tau_i}(\sigma,\sigma') \leq b_i(h+1). \]
Because we have shown this inequation to hold for each $i \in B$, we conclude that
\begin{equation}
 \bigwedge_{i \in B} \expsem{\tau_i}(\sigma,\sigma') \leq b_i(h+1). \label{Eq:Final}
\end{equation}
Recall that $\sigma$ and $\sigma'$ are the initial and final states of $\taug$,
and that, by the construction of $\taug$, these are also the initial and final
states of the original trace $t$ of $P$.  But $t$ was an arbitrary feasible
execution trace of $P$ of height up to $h+1$, and therefore we conclude that
\eqref{Final} holds if $(\sigma,\sigma')$ are the initial and final states of
any feasible execution trace of $P$ of height up to $h+1$.  Thus,
\[ \bigwedge_{i \in B} \forall v \in V_{\tau_i}(P,h+1). v \leq b_i(h+1), \]
and the proof is complete.
\end{proof}


}

\end{document}